\documentstyle[prl,aps,floats,psfig]{revtex}

\begin{document}
\draft

\flushbottom
\twocolumn[\hsize\textwidth\columnwidth\hsize
\csname @twocolumnfalse\endcsname

\title{Dynamic generation of maximally entangled photon multiplets by 
adiabatic passage}
\author{W. Lange\footnotemark\ and H. J. Kimble}
\address{Norman Bridge Laboratory of Physics 12-33,
California Institute of Technology, Pasadena, California 91125}
\date{10 August, 1999}

\maketitle
\begin{abstract}
The adiabatic passage scheme for quantum state synthesis, in which atomic
Zeeman coherences are mapped to photon states in an optical cavity, is extended
to the general case of two degenerate cavity modes with orthogonal
polarization. Analytical calculations of the dressed-state structure and Monte
Carlo wave-function simulations of the system dynamics show that, for a suitably
chosen cavity detuning, it is possible to generate states of photon multiplets
that are maximally entangled in polarization. These states display
nonclassical correlations of the type described by Greenberger, Horne, and
Zeilinger (GHZ). An experimental scheme to realize a GHZ measurement using
coincidence detection of the photons escaping from the cavity is proposed. The
correlations are found to originate in the dynamics of the adiabatic passage
and persist even if cavity decay and GHZ state synthesis compete on the same
time scale. Beyond entangled field states, it is also possible to generate
entanglement between photons and the atom by using a different atomic
transition and initial Zeeman state.
\end{abstract}

\vskip2pc]

\footnotetext{\footnotemark Present address:
Max-Planck-Institut f\"ur Quantenoptik,\\ 
D-85748 Garching, Germany.\\
Electronic address: wfl@mpq.mpg.de}

\section{Introduction}

A quantum system consisting of several components is called entangled if its
state cannot be expressed as a direct product of substates for the
constituents. Such a system has no classical analog and hence shows distinct
quantum mechanical features. Of particular interest are multiplets of particles
in maximally entangled states, in which a measurement of the energy eigenvalue
of one particle completely determines the outcome of the same measurement for
all the others. An example in the case of two particles are the EPR or
Bell states \cite{Bell65}, which give the largest contradiction with the local
realism assumption of Einstein, Podolsky, and Rosen \cite{EPR35}. Greenberger,
Horne, and Zeilinger \cite{GHZ89,Green90} and later Mermin \cite{Mermi90} have
shown that a much stronger and more direct test of local realism may be
achieved by using three or more maximally entangled particles. Violation of
local realism can then be obtained in a single set of experiments, whereas EPR
correlations between two particles show up only statistically in a series of
measurements. 

A number of different experimental schemes have been proposed to generate a
maximally entangled or GHZ state involving three or more photons or atoms
\cite{GHZ89}. Proposals for optical experiments typically use photon pairs
generated in parametric down conversion to obtain GHZ correlations
\cite{Zukow91,Reid92,Klysh93} or polarization correlations among photons
emitted in cascaded atomic decays \cite{Green90}. Other schemes suggest
utilizing the QED interaction of atoms and field modes of a cavity to generate
entanglement. Three or four two-level Rydberg atoms may be prepared in an
entangled state through their interaction with the field in a microwave cavity
\cite{Cirac94,Haroc95,Gerry96a,Gerry96b}. Conversely, three microwave cavities
may be entangled due to the interaction with a Rydberg atom \cite{Gerry96}. Yet
another suggestion uses a single photon in a double Mach-Zehnder interferometer
to create a three-atom GHZ state \cite{Hardy91b}. There are also mixed schemes
with entanglement between different types of subsystems, for example four
cavity modes and a single atom \cite{Wodki93}. While in most proposals the
entangled particles originate in a single localized source, GHZ correlations
may also be produced using particles from separate sources
\cite{Yurke92a,Zeili97}. In this way recently photon states were realized
that retrodictively reproduce GHZ statistics of photocounts \cite{Bouwm99}.
Another system that should be capable of producing entanglement are two-level
ions in a linear quadrupole trap, subjected to appropriately timed pulses
\cite{Winel92}. Some of these schemes allow the subsystems to be probed in
spatially separated locations, which is important if the issues of
nonseparability and nonlocality in quantum mechanics are to be investigated. In
other cases such a separation of the subsystems is impossible in principle,
thus precluding tests of the causal dependence between the measurements. In
spite of the many theoretical proposals, however, to date no experimental
realization of a GHZ state with three or more particles has been achieved.

Here, we present a novel approach to the generation of GHZ correlations,
adapting a method from the field of quantum-state synthesis in cavity QED
\cite{Parki93,Vogel93,Law96a}. The particles we consider are photons in a
single longitudinal mode of an optical resonator. One way to create ``arbitrary"
nonclassical states of the field in such a system is the method of adiabatic
passage \cite{Parki93,Parki95,Kuhn99}. In this protocol, an atom that is prepared 
in a coherent superposition of Zeeman ground states sequentially traverses a
resonant quantized cavity mode and a coherent-state laser field with
overlapping spatial profiles. If the transit is slow enough (adiabatic), the
system remains in a single quantum state and the coherence of the Zeeman
sublevels is mapped to the photon distribution of the cavity mode. Therefore,
the number of quantum states of the field that may be synthesized is only
limited by the possible Zeeman state superpositions of the atom. Previous work
\cite{Parki93,Parki95} shows this scheme to be robust, without the requirement
for precise control of pulse area or timing.

In our scheme for the generation of a GHZ state, the cavity photons are
entangled in polarization. Since in the previous theory of adiabatic passage
\cite{Parki93,Parki95} only a single polarization was considered (which was
taken to be $\sigma_-$, implying an infinitely detuned $\sigma_+$-polarized
mode), we start in Sec.~\ref{Sec:Model} by extending the method to the case of
two degenerate orthogonal polarizations for the cavity mode. The model allows
us to find a straightforward procedure for the synthesis of a GHZ state of
polarization entangled photons and to calculate the efficiency of its
generation. Throughout the paper we utilize parameters already realized in
present day cavity QED experiments. Note that from an experimental point of
view, our results are relevant even for the generation of Fock states of a {\em
single} polarization, since in realistic Fabry-P{\'e}rot resonators, modes with
both transverse polarizations are always present. Although in practice small
biases towards one polarization may occur (e.g., due to birefringence of the
mirrors), our model assumes that the frequency splitting of the polarization
modes of the cavity is negligible (degenerate case). However, in
Sec.~\ref{Sec:QMC} we present numerical calculations for finite mode splitting
(in the circular polarization basis) to derive a quantitative criterion for the
validity of the degenerate model. These results also show how large a mode
splitting is necessary for a single transverse mode description of the system
to be applicable. 

In the single polarization case \cite{Parki93,Parki95}, the adiabatic dynamics
evolves in a one-dimensional space along the so-called {\em dark state}, a
superposition of ground states which is decoupled from excited atomic levels.
Each atomic ground state within the manifold of Zeeman substates is correlated
with one (and only one) photon number of the cavity field, as any given Zeeman
state is reached by a unique sequence of photon absorption and emission cycles.
Thus a shift of the atomic population is always accompanied by a corresponding
shift of the photon distribution. By contrast, with both (degenerate)
polarizations of the cavity field taken into account, the system behavior is
considerably more complex. Population may be shifted to higher or lower
magnetic sublevels, depending on the polarization of the photons involved. As a
consequence there is no intrinsic limit to the number of photons that may be
generated by adiabatic passage. The resulting final states of the cavity field
are no longer expected to be pure Fock states, but a complex superposition of
number states for both polarizations, even when the initial atomic state is a
single Zeeman substate. This is shown in Sec.~\ref{Sec:States} by analyzing the
structure of the relevant eigenstates.

After characterizing the dynamics of state synthesis by adiabatic passage with
two polarizations, we proceed by proposing a new strategy to arrive at pure
Fock states of the cavity field by making use of a suitable detuning of the
cavity field. The feasibility of the method is demonstrated in
Sec.~\ref{Sec:QMC} by a quantum Monte Carlo simulation of the system dynamics.
An important result is that there is no restriction on our method with respect
to polarization: the scheme is capable of generating number states of either
cavity polarization or even arbitrary superpositions of both. The last fact
opens the door for the generation of a GHZ state of polarization entangled
photons starting from a suitable superposition of Zeeman ground states. Details
of this new protocol are explained in Sec.~\ref{Sec:GHZprep}. 

In a realistic experiment, the synthesis of a GHZ state in a cavity will
certainly be affected by energy dissipation. In Sec.~\ref{Sec:GHZdet} we
incorporate this fact into the model and utilize it to detect GHZ correlations
by coincidence measurements of the photons escaping from a leaky cavity.
Somewhat surprisingly, we find that even when decay sets in before the
generation of the GHZ state is finished (i.e., when the cavity decay and
adiabatic passage times are comparable), correlations between the photons
persist. Obviously, these correlations originate in the dynamics of the
adiabatic process rather than in the properties of a stationary final state of
the system.

A variation of the method that creates entanglement among photons and atoms is
presented in Sec.~\ref{Sec:GHZAtom}. Here, the initial state for the adiabatic
passage is a single Zeeman ground state, that may be easily prepared by
optical pumping with $\pi$-polarized light. GHZ correlations are detected by a
coincidence measurement of the photons emitted from the cavity and information
on the atomic state obtained from a ``Stern-Gerlach" device. Our results are
summarized in Sec.~\ref{Sec:Concl}.

\section{Model of the atom-cavity system}
\label{Sec:Model}

\begin{figure}[tb]
\vbox{
\centerline{\psfig{file=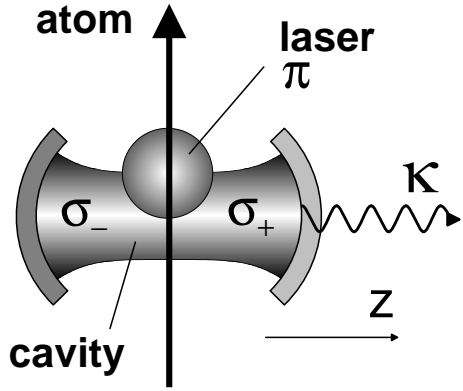,width=0.7\hsize}}\par\vspace{5mm}
\caption{Setup of the two-polarization adiabatic passage scheme.
With spatial quantization along the $z$ axis, the cavity supports
a $\sigma^+$- and a degenerate $\sigma^-$-polarized mode. The atomic beam
successively traverses the cavity waist and the $\pi$-polarized
pump laser beam, which is oriented perpendicular to the direction
of the atoms and the cavity axis.}
\label{Fig:Setup}}
\end{figure}

The experimental configuration we consider is similar to the one presented in
\cite{Parki93} and is depicted in Fig.~\ref{Fig:Setup}. A moving atom
(transition frequency $\omega_a$) first encounters the waist of a quantized
cavity field and subsequently a pump-laser field injected perpendicular to both
cavity axis and the atom's direction of propagation. Cavity and pump fields
overlap partially, so that there is a region where in the rest frame of the
atom the pump field increases as the cavity field decreases. The cavity field
is described by two transverse-polarization states of a single longitudinal
mode. With the cavity axis as the direction for spatial quantization, the
polarizations are taken to be $\sigma^{+}$ and $\sigma^{-}$ and the resonance
frequencies of the corresponding modes denoted as $\omega_{c\pm}$. The ability
to accomodate arbitrary polarization of the cavity field distinguishes the
model presented here from that of Refs.~\cite{Parki93,Parki95}. It not only
allows us to investigate polarization entangled states of the cavity field,
such as the maximally entangled GHZ states, but also leads to an improved
scheme for the generation of single-polarization Fock states in a realistic
experimental situation.

%%%%%%%%%%%%%%%%%%%% Fig 1 %%%%%%%%%%%%%%%%%%%%%%%%%%%%%%%%%%%%%%%
%%%%%%%%%%%%%%%%%%%% Fig 1 %%%%%%%%%%%%%%%%%%%%%%%%%%%%%%%%%%%%%%%

We assume here that the atomic levels involved are the Zeeman substates of
ground and excited levels $F_{g}$ and $F_{e}$, although more general
configurations can also be employed \cite{Law96a,Law97,Cirac97}. In most of the
paper we concentrate on the case $F_g=F_e$, where the transition $(m_g=0)
\rightarrow (m_e=0)$ is electric dipole forbidden, so that a $\pi$-polarized
pump beam does not couple to atoms in the $m_g=0$ state of the lower Zeeman
manifold. Therefore, if by a suitable interaction with pump field and cavity
mode, atomic population is pushed towards levels with decreasing $|m_g|$, the
process will terminate once the level $(m_g=0)$ is reached, independent of the
initial atomic distribution. Note that the coupling coefficients among the
various Zeeman states for the $F_{g}\leftrightarrow F_{e}$ transition are
symmetric under a change of sign of the magnetic quantum number, which is
important for the generation of states involving photons of different
polarization.

\begin{figure}[tb]
\vbox{
\centerline{\psfig{file=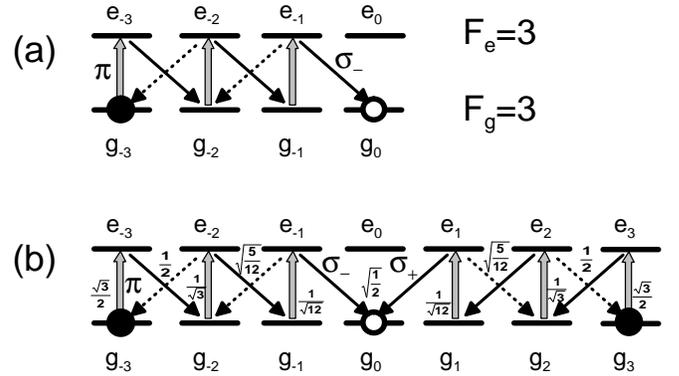,width=\hsize}}\par\vspace{5mm}
\caption{Scheme of the $(F_g=3)$ and $(F_e=3)$ hyperfine levels,
showing the transitions relevant for adiabatic passage starting
from two different initial preparations of Zeeman
sublevels (indicated by the solid circles): (a) $m_g=-3$;
(b) superposition of $m_g=-3$ and $m_g=+3$. The wide arrows 
symbolize the $\pi$-polarized (classical) pump transitions. 
The predominant cavity-induced transition from a given excited 
state {\sf e}$_m$ [larger Clebsch-Gordan coefficient as indicated 
by the numbers in diagram (b)] is drawn as a solid, the weaker
emission into the orthogonal polarization mode as a dotted arrow.
Note that at the end of the passage the atom is always found in
the $m_g=0$ level (open circle), even in the presence of two 
cavity polarizations.}
\label{Fig:Csscheme}}
\end{figure}

The maximum number of photons that can be generated with the scheme is achieved
when the atom is prepared in the outer Zeeman ground states $(m_g=\pm F_g)$ or
a coherent superposition of both. As shown below, ideal adiabatic passage in
this case will generate $F_g$ (entangled) photons distributed among both cavity
modes. With cesium (nuclear spin $I=7/2, F_g=3$ or $4$) as an example, states
with three or four entangled photons are therefore possible.
Figure~\ref{Fig:Csscheme} shows the relevant pump and cavity-induced
transitions in the case $F_g=F_e=3$ for two possible initial preparations of
the Zeeman sublevels.

%%%%%%%%%%%%%%%%%%%% Fig 2 %%%%%%%%%%%%%%%%%%%%%%%%%%%%%%%%%%%%%%%
%%%%%%%%%%%%%%%%%%%% Fig 2 %%%%%%%%%%%%%%%%%%%%%%%%%%%%%%%%%%%%%%%

In our model, the cavity field is described by annihilation and creation
operators $a_+,a_+^\dagger$ for the $\sigma_+$-polarized mode and
$a_-,a_-^\dagger$ for $\sigma_-$ polarization, while the pump field is treated
classically (i.e., as a coherent state in a regime of weak coupling). The
atomic degrees of freedom are described by operators $A_\sigma$, which are
composed of a sum over spin-$1/2$ operators for the Zeeman levels
$|F_g,m_g\rangle$ and $|F_e,m_e\rangle$ with weights corresponding to the
Clebsch-Gordan coefficients for dipole coupling with polarization $\sigma$,
namely
\begin{equation}
A_\sigma=\!\!\sum\limits_{m_g,m_e}\! 
|F_g m_g\rangle\langle F_g m_g; 1 \,\sigma | F_e m_e \rangle\langle F_e m_e |,
\; \sigma=0,\pm 1.
\end{equation}
As the atom moves across the cavity, it undergoes a time-dependent interaction
with the fields, which is expressed by a variable coupling constant $g(t)$ for
the two quantized cavity modes and a Rabi frequency $\Omega(t)$ for the
classical pump field. The time scale is determined by the atomic velocity.
Alternative time-dependent coupling schemes for the case of a stationary atom
located in the cavity field \cite{Law96a,Law97,Cirac97} may be treated with the
same formalism.

With the above definitions, the Hamiltonian for the atom and the two fields in
a frame rotating at the frequency $\omega_l$ of the classical field $\Omega(t)$
is given by
\begin{eqnarray}
H_{\text{int}}(t)&=&
\hbar\delta_+ a_+^\dagger a_+ + \hbar\delta_- a_-^\dagger a_-
\nonumber\\
&&
-i\hbar g(t) \left(a_+^\dagger A_{+1} - A_{+1}^\dagger a_+ \right)
\nonumber\\
&&
-i\hbar g(t) \left(a_-^\dagger A_{-1} - A_{-1}^\dagger a_- \right)
\nonumber\\
&&
+i\hbar \Omega(t) \left(A_0 - A_0^\dagger \right).
\label{Hamilt}
\end{eqnarray}
Here, $\delta_\pm=\omega_{c\pm}-\omega_l$ is the detuning between the 
cavity frequency $\omega_{c\pm}$ for the $\sigma_\pm$-polarized mode and 
the classical field, which is assumed to be at resonance with the atomic
transition ($\omega_a=\omega_l$), throughout the present analysis.

A basis for the Hilbert space of the atom-cavity system is 
provided by tensor products of the atomic states and the number
states for each polarization of the cavity mode. We use the notation 
\begin{equation}
|x_m,n_+,n_-\rangle = |x_m\rangle \otimes |n_+\rangle \otimes |n_-\rangle,
\label{Basis}
\end{equation}
with $x=e,g$ specifying excited
and ground state, $m=-F_x,\dots, F_x$ the number of the magnetic sublevel,
and $n_\pm$ the photon number in the respective cavity mode. 

In this paper the initial preparation of the system is not described
explicitly. We assume that before an atom enters the interaction zone both
modes of the cavity field are in the vacuum state and the atom is prepared in
an arbitrary coherent superposition of Zeeman ground states. This may be
achieved by radio-frequency pumping \cite{Carte95} or a series of optical Raman
pulses \cite{Law98}.

In our initial discussion in the following section, atomic spontaneous
emission and cavity decay will be neglected. They will be included in
Secs.~\ref{Sec:QMC} and~\ref{Sec:GHZprep} in a master equation approach
using the quantum Monte Carlo wave-function technique.

\section{Dressed states and dark space}
\label{Sec:States}

\begin{figure}[tb]
\vbox{
\centerline{\psfig{file=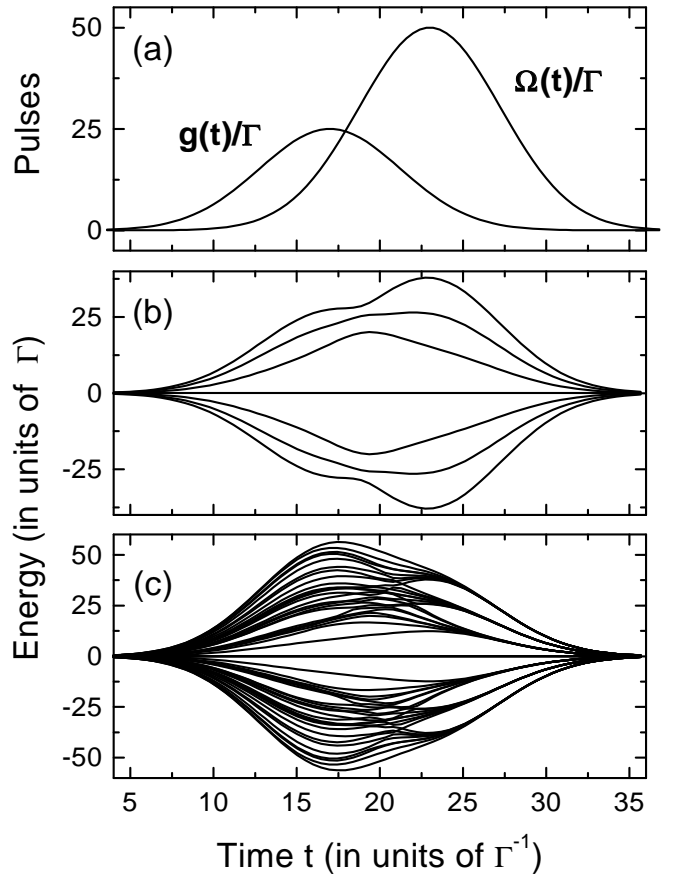,width=\hsize}}\par\vspace{5mm}
\caption{(a) Time evolution of the coupling $g(t)$ and Rabi frequency
$\Omega(t)$ of the pump field, which are assumed to be Gaussian pulses 
with FWHM $10\Gamma^{-1}$; they are centered at $t=17\Gamma^{-1}$ and 
$t=23\Gamma^{-1}$, respectively, and have amplitudes $g_0=25\Gamma$ and 
$\Omega_0=50\Gamma$;
(b) energy eigenvalues of $H_{\text{int}}$ in the single-polarization 
case; (c) energy eigenvalues for two polarizations.
Atom, cavity modes, and pump field are at resonance 
($\omega_a=\omega_{c+}=\omega_{c-}=\omega_l$).}
\label{Fig:EV}}
\end{figure}

Fundamental aspects of the system evolution in the two-polarization
adiabatic passage scheme may be inferred from the structure of the
eigenvalues and eigenvectors of $H_{\text{int}}(t)$ during the transit
of a single atom. As a specific example we consider the
$(6S_{1/2},F_g=3)\rightarrow (6P_{1/2},F_e=3)$ transition in cesium
(cf.\ Fig.~\ref{Fig:Csscheme}) at $852.36$ nm. For the system parameters
we have chosen values reached in recent cavity QED experiments with high
finesse optical resonators
\cite{ICAP94,Turch95,Turch95b,Mabuc96,Hood98}. In particular, a peak
atom-cavity coupling $g_0=25 \Gamma$ on the resonator axis was used with
$\Gamma/2\pi=5\;\text{MHz}$ being the spontaneous emission rate for the
cesium D2 line (i.e., $\Gamma^{-1}$ is the population lifetime). Both
cavity modes and the pump beam are assumed to have a Gaussian transverse
shape with a full width at half maximum (FWHM) $w=10\Gamma^{-1}$, but
the pump beam has its center displaced along the atomic beam by an
amount of $0.6 w$ relative to the cavity mode [Fig.~\ref{Fig:EV}(a)]. 
The maximum Rabi
frequency $\Omega_0$ of the pump field is chosen as twice the coupling
constant $g_0$ (i.e., $\Omega_0=50 \Gamma$ for the above specified
coupling). These values were found to yield optimum results in the
single polarization case \cite{Parki93,Parki95} as well as in the
present work. Small variations in the shape and intensity of the field
distributions on the order of a few percent, however, do not lead to
substantially different results.

%%%%%%%%%%%%%%%%%%%% Fig 3 %%%%%%%%%%%%%%%%%%%%%%%%%%%%%%%%%%%%%%%
%%%%%%%%%%%%%%%%%%%% Fig 3 %%%%%%%%%%%%%%%%%%%%%%%%%%%%%%%%%%%%%%%

With the two cavity modes in the vacuum state and 
the atom prepared in the lowest Zeeman ground state ($m_g=-3$),
the initial wave function in the notation of Eq.~(\ref{Basis}) is
\begin{equation}
|\Psi_{\text{ini}}\rangle=|g_{-3},0,0\rangle.
\end{equation}
The subsequent analysis will be restricted to the manifold
of states that are coupled to $|\Psi_{\text{ini}}\rangle$ by the 
interaction Hamiltonian $H_{\text{int}}$.

In order to compare the system behavior for the cases of one and two
polarizations, we first calculate the eigenvalues of $H_{\text{int}}$
when only the $\sigma_-$ mode interacts with the atom, corresponding to
the single polarization analysis of Refs.~\cite{Parki93,Parki95}. To
this end the terms involving $a_+$-operators are omitted from the
Hamiltonian~(\ref{Hamilt}). Figure~\ref{Fig:EV}(b) shows the eigenvalues
obtained as a function of transit time for typical experimental
parameters and a cavity detuning $\delta_-=0$. There are seven
nondegenerate energy levels that are also referred to as dressed
states since they describe the energy of atom and the surrounding
fields. The dressed state employed for the adiabatic passage is the one
that does not couple to excited atomic states and hence does not
experience an ac-Stark shift due to either the cavity or pump fields
(dark state). Its energy eigenvalue is $E=0$ and the corresponding
eigenvector in the basis~(\ref{Basis}), up to a normalization factor, is
\begin{eqnarray}
|E_0\rangle &=&
-\sqrt{15} g^3(t)	 	|g_{-3},0,0\rangle
+\sqrt{45} g^2(t) \Omega(t) 	|g_{-2},0,1\rangle
\nonumber\\&&
-\sqrt{18} g(t) \Omega^2(t) 	|g_{-1},0,2\rangle
+	    \Omega^3(t) 	|g_{0},0,3\rangle.
\label{DarkState}
\end{eqnarray}

This state has the property that if the system is prepared in $
|g_{-3},0,0\rangle$ in a region of $\Omega\approx 0$, then slowly
increasing $\Omega(t)$ while reducing $g(t)$ to zero ($g/\Omega\approx
0$) will adiabatically transform the initial state to
$|g_{0},0,3\rangle$, i.e., a state with three $\sigma_-$ photons in the
cavity mode. As is apparent from Eq.~(\ref{DarkState}), the creation of
each photon is accompanied by an increase in the magnetic quantum number
$m_g$ of the atom. Since for a $\pi$-polarized classical laser field no
more pumping can occur once the state $m_g=0$ is reached, exactly three
photons will be found in the cavity mode at the end of the adiabatic
passage. This method has been analyzed as a scheme for the generation of
Fock states and for more general superposition states of the field
arising from superpositions of initial Zeeman states in the $F_{g}$
mainifold \cite{Parki93,Parki95}.

These conclusions are no longer valid if a second, orthogonally
polarized cavity mode near resonance is taken into account, as a
calculation of the eigenvalues of $H_{\text{int}}$ for two cavity
polarizations with equal frequencies demonstrates. According to
Fig.~\ref{Fig:EV}(c), the resulting level structure is considerably more
complex than in the single polarization case, with a large number of
levels undergoing crossings and anticrossings. This is due to the fact
that in contrast to the single-polarization case there is no intrinsic
limit to the number of photons that may be deposited in the cavity mode,
as will be explained below. The numerical simulations
(Sec.~\ref{Sec:QMC}) show that not all the levels accessible to the
system contribute equally to the dynamics. For example, a cutoff of six
photons in both cavity modes was found to produce sufficiently accurate
results and was used in Fig.~\ref{Fig:EV}(c).

Another important implication of the second polarization is that the
dark state at the energy $E=0$ becomes highly degenerate, the degree of
degeneracy being limited only by the cutoff photon number. There is no
longer a single dark state but rather a {\em dark space} in which the
dynamics evolve in the adiabatic case. Consequently, no simple final
state of the atom-cavity system is expected to be generated by the
adiabatic passage, but rather a complex superposition state in the dark
subspace. 

%%%%%%%%%%%%%%%%%%%% Fig 4 %%%%%%%%%%%%%%%%%%%%%%%%%%%%%%%%%%%%%%%
\begin{figure}[tb]
\vbox{
\centerline{\psfig{file=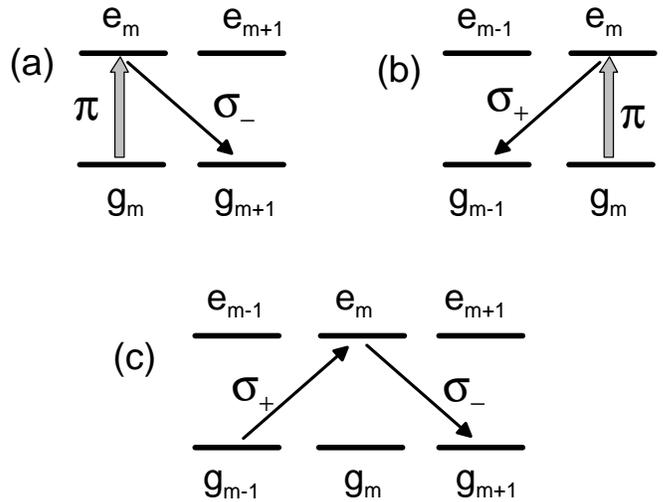,width=\hsize}}\par\vspace{5mm}
\caption{Fundamental processes of the dynamics within the dark space.
(a) Scattering of a pump photon into the $\sigma_-$ cavity mode;
(b) scattering of a pump photon into the $\sigma_+$-cavity mode;
(c) polarization flip of a cavity photon.}
\label{Fig:Processes}}
\end{figure}
%%%%%%%%%%%%%%%%%%%% Fig 4 %%%%%%%%%%%%%%%%%%%%%%%%%%%%%%%%%%%%%%%

In the single polarization case only one fundamental process is involved
in the evolution of the dark state~(\ref{DarkState}), in which a photon
from the pump field is scattered into the $\sigma_-$ cavity mode,
increasing the magnetic quantum number by one
[Fig.~\ref{Fig:Processes}(a)]. The process repeats until the atom
reaches the Zeeman ground state with $m_g=0$. As will be analyzed in
more detail below, with two polarizations present the final state of the
adiabatic passage still has the atom in $m_g=0$. However, two additional
processes contribute to the dynamics within the dark eigenspace of
energy $E=0$. In one process, a linearly polarized photon from the pump
field is scattered into the $\sigma_+$-polarized cavity mode,
accompanied by a decrease in the magnetic quantum number of the atom
[Fig.~\ref{Fig:Processes}(b)]. As a secondary consequence, the system
has to go through an additional cycle of the type shown in
Fig.~\ref{Fig:Processes}(a) to reach the final atomic ground state
$m_g=0$. Effectively the system has gone through a loop in the atomic
level scheme, creating a pair of $\sigma_+$/$\sigma_-$ photons in the
process. Depending on the system parameters, several of these loops may
occur. Clearly, loop processes must be avoided if a pure number state is
to be generated, as they lead to superposition states of different (and
somewhat uncontrollable) photon numbers in the cavity.

The second additional process is the destruction of a $\sigma_+$ photon
followed by the creation of a $\sigma_-$ photon (and conversely) with no
pump photons involved [Fig.~\ref{Fig:Processes}(c)]. In this case the
photon number in the cavity is not changed, but the polarization of one
photon is flipped. At the same time the atom changes its Zeeman substate
by two. Due to this process, states beyond the $m_g=0$ Zeeman ground
state may become temporarily occupied, which is impossible in the single
polarization case. The branching ratio determining the relative
contribution of each of the three processes of Fig.~\ref{Fig:Processes}
(at resonance) depends on three factors: the occupation of the cavity
mode (influencing induced emission), the Clebsch-Gordan coefficients of
the relevant transitions (cf.\ Fig.~\ref{Fig:Csscheme}), and the temporal
structure of the couplings and fields. By preparing the atom in the
lowest Zeeman state it is guaranteed that the first transition is of
type (a), enhancing the probability for further emissions with the
desired $\sigma_-$ polarization. The Clebsch-Gordan coefficients also
favor transitions towards decreasing $|m_g|$ so that the adiabatic
passage method with two polarizations is still expected to work, albeit
with an efficiency degraded by the processes of
Figs.~\ref{Fig:Processes}(b,c). To obtain pure Fock states of the cavity
field for a given initial Zeeman substate, a method to suppress these
undesired transitions must be found. We will demonstrate in the
following that a finite detuning of the cavity from resonance
effectively achieves this end.

To analyze the structure of the dark space manifold, it is convenient to
parametrize the vectors spanning the $E=0$ eigenspace by the number of
photons contained. As photons are transferred to the cavity during
adiabatic passage, the initial occupation of the modes (when the pump
field $\Omega=0$ and the atom is in the state $g_{-3}$) is used to
characterize each eigenstate unambiguously. Only such states with an
even initial photon number contribute, corresponding to the fact that
nonadiabatic coupling to states with higher photon number only occurs
by the loop process described above (which adds two photons to the
system). For example, for the initial atomic state $|g_{-3}\rangle$,
explicit expressions for the eigenstates corresponding to total initial
photon numbers zero, two, and four are given by
\begin{mathletters}
\label{eigenstates}
\begin{eqnarray}
%|E_{0}&&\rangle  =  {\cal{N}}_0\;
|E_{0}\rangle  &=& {\cal{N}}_0\; 
[
-\sqrt{15} g^3(t)	 	|g_{-3},0,0\rangle \nonumber\\
&& +\sqrt{45} g^2(t) \Omega(t) 	|g_{-2},0,1\rangle \nonumber\\
&& -\sqrt{18} g(t) \Omega^2(t) 	|g_{-1},0,2\rangle
+	    \Omega^3(t) 	|g_{0},0,3\rangle]\\
%|E_{1}&&\rangle  =  {\cal{N}}_1\;\nonumber\\
|E_{1}\rangle  &=&  {\cal{N}}_1\;
[\sqrt{60} g^3(t)	 	|g_{-3},1,1\rangle\nonumber\\
&&-\sqrt{90} g^2(t) \Omega(t) 	|g_{-2},1,2\rangle\nonumber\\
&&+\sqrt{24} g(t) \Omega^2(t) 	|g_{-1},1,3\rangle
-		\Omega^3(t) 	|g_{0},1,4\rangle\nonumber\\
&&  +\sqrt{18} g^3(t)	 	|g_{-1},0,2\rangle
-\sqrt{36} g^2(t) \Omega(t) 	|g_{0},0,3\rangle \nonumber\\
&&  +\sqrt{6}  g(t) \Omega^2(t) 	|g_{+1},0,4\rangle]\\
%|E_{2}&&\rangle =  {\cal{N}}_2\; \nonumber\\
|E_{2}\rangle &=&  {\cal{N}}_2\;
%&& 
[ -\sqrt{150} g^3(t)      	|g_{-3},2,2\rangle\nonumber\\
&&+\sqrt{150} g^2(t) \Omega(t) 	|g_{-2},2,3\rangle\nonumber\\
&&-\sqrt{30}  g(t) \Omega^2(t) 	|g_{-1},2,4\rangle
+		 \Omega^3(t) 	|g_{0},2,5\rangle\nonumber\\
&&-\sqrt{60}  g^3(t)      	|g_{-1},1,3\rangle
+\sqrt{90}  g^2(t) \Omega(t) 	|g_{0},1,4\rangle\nonumber\\
&& -\sqrt{12}  g(t) \Omega^2(t) 	|g_{+1},1,5\rangle\nonumber\\
&& -\sqrt{15}  g(t)^3		|g_{+1},0,4\rangle\nonumber\\
&&+\sqrt{15}  g(t)^2\Omega(t) 	|g_{+2},0,5\rangle ]
\end{eqnarray}
\end{mathletters}%
with the symbols ${\cal{N}}_k$ denoting normalization constants.  
Note that $|E_0\rangle$ is identical to the single polarization 
dark state of Eq.~(\ref{DarkState}).

%%%%%%%%%%%%%%%%%%%% Fig 5 %%%%%%%%%%%%%%%%%%%%%%%%%%%%%%%%%%%%%%%
\begin{figure}[tb]
\vbox{
\centerline{\psfig{file=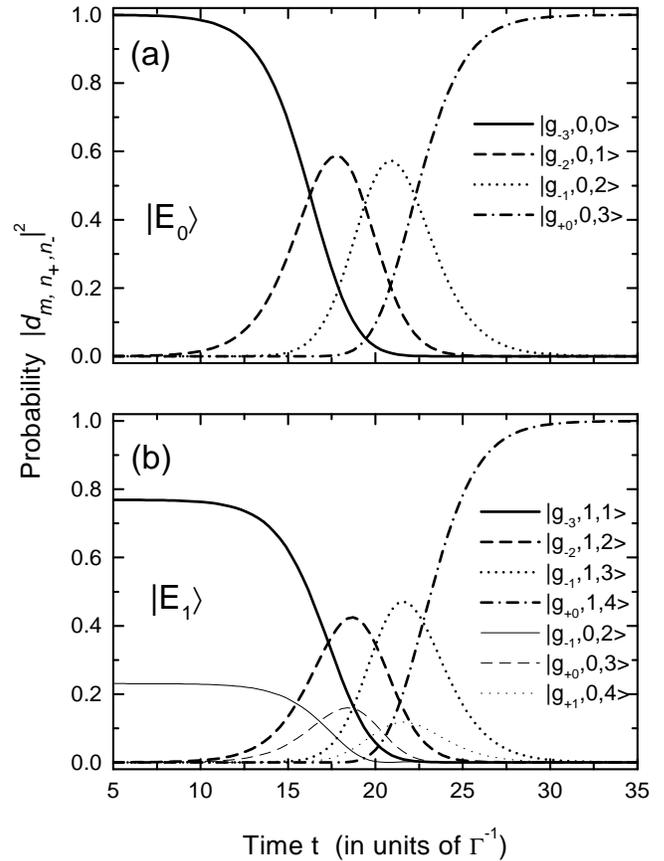,width=\hsize}}\par\vspace{5mm}
\caption{\sloppy Time evolution of the first two eigenvectors with degenerate 
eigenvalues ($E=0$) in terms of the basis state coefficients 
$\left|d_{m,n_+,n_-}\right|^2$. They are defined by the expansion 
$|E_k(t)\rangle=$ $\sum_{m,n\pm} d_{m,n_+,n_-}^{(k)}(t) |g_m,n_+,n_-\rangle$
[cf.\ Eq.~(\ref{eigenstates})].
(a) Eigenvector with no $\sigma_+$ contribution, corresponding to 
the dark state of the single-polarization case; 
(b) eigenvector with one $\sigma_+$ photon in the cavity; note the
additional contributions obtained by polarization flips from 
$\sigma_+$ to $\sigma_-$ (thin lines).}
\label{Fig:Eigenstates}}
\end{figure}
%%%%%%%%%%%%%%%%%%%% Fig 5 %%%%%%%%%%%%%%%%%%%%%%%%%%%%%%%%%%%%%%%

The evolution of the first two eigenstates ($|E_0\rangle$ and
$|E_1\rangle$) in terms of the basis state components
$|g_m,n_+,n_-\rangle$ is plotted in Fig.~\ref{Fig:Eigenstates}. As the
adiabatic passage proceeds, each state acquires exactly three additional
$\sigma_-$ photons due to the process of Fig.~\ref{Fig:Processes}(a).
Moreover, occupation is redistributed by the polarization flips shown in
Fig.~\ref{Fig:Processes}(c). For a change of eigenstate to occur, the
system must undergo a loop process involving the generation of a
$\sigma_+/\sigma_-$ photon pair. These latter events are the reason for
the failure of standard adiabatic passage to produce a pure Fock state
in the two-polarization case with degenerate frequencies.

Possible final states of the system may be deduced from
Eqs.~(\ref{eigenstates}) by taking the limit $g(t)\rightarrow 0$. As
expected from the vanishing coupling of the pump field in the central
Zeeman state, the atom emerges from the adiabatic passage always in the
$m=0$ ground state as in the single polarization case. This is an
important fact, because it ensures that the atom is not entangled with
the field. The cavity modes are found in a superposition of the states
$|0,3\rangle$, $|1,4\rangle$, $|2,5\rangle, \dots$, so that the general
expression for the final state of the system is
\begin{equation}
\label{finalstate}
|g_0\rangle\;\otimes\;\sum\limits_{k=0}^\infty c_k |k,3+k\rangle.
\end{equation}
Here, the notation $|n_+,n_-\rangle$ is used to denote number
states of the two-mode cavity field.

As a consequence of Eq.~(\ref{finalstate}), along with the desired
$\sigma_-$ Fock state $|0,3\rangle$, there is a finite probability for
additional pairs of $\sigma_+$  and $\sigma_-$ photons in the cavity
field. The size of those contributions, i.e., the modulus of the
coefficients $c_k$ in Eq.~(\ref{finalstate}), must be determined from
the dynamics generated by $H_{\text{int}}$ (see Sec.~\ref{Sec:QMC}).
However, an estimate for $|c_1|^2$ can be obtained by restricting the
system to the subspace spanned by the two lowest eigenvectors of the
dark space, with $|E_0\rangle$ being the state initially occupied and
$|E_1\rangle$ the first state with a $\sigma_+$ polarization
contribution. As the eigenvectors in Eqs.~(\ref{eigenstates}) are time
dependent, diabatic Landau-Zener-type transitions between the associated
eigenstates occur. The transition probability is obtained by integrating
the Schr\"odinger equation for a two-level system with time-dependent
basis states:
\begin{equation}
\label{twolevelSE}
\left( i \hbar {\partial\over\partial t} - H_{\text{int}} \right)
\left[ c_0(t)\; |E_0(t)\rangle\; +\; c_1(t)\; |E_1(t)\rangle \right] = 0.
\end{equation}
Starting in state $|E_0\rangle$ at $t\rightarrow -\infty$, 
we derive the
probability $|c_1|^2$ for a diabatic transition to state $|E_1\rangle$ at 
$t\rightarrow \infty$ to be given by the following expression:
\begin{equation}
\label{LandauZener}
|c_1|^2\!=\!\!\!\left(\sin\left|
\int\limits_{-\infty}^{+\infty}\!\! %\;
{{\left\langle E_1\left|{\partial\over\partial t}\right|E_0\right\rangle}%
\over{\sqrt{1-|\langle E_1|E_0\rangle|^2}}}%\; %
e^{-i [E_1(\tau)-E_0(\tau)]\tau/\hbar}%\;  
d\tau \right|
\right)^{\!\!\!\! 2}.
\end{equation}

From Eqs.~(\ref{eigenstates}) one finds that the matrix elements
appearing under the integral do not vanish, as the eigenvectors and
their derivatives are not orthogonal [because of the assumed classical
field for $\Omega(t)$]. The size of the relevant scalar products as a
function of time and hence the diabatic transition probability peak 
in the middle of the atomic transit. For the parameters of 
Fig.~\ref{Fig:EV}, there is a 22\% probability for the system to
undergo a Landau-Zener transition during the adiabatic passage and end
up in the state $|g_0,1,4\rangle$. The model of two degenerate dark
states thus demonstrates that adiabatic passage as described in
\cite{Parki93} is not well suited to generate number states in a cavity
if the two polarizations of the cavity field are closely spaced in
frequency.

\begin{figure}[tb]
\vbox{
\centerline{\psfig{file=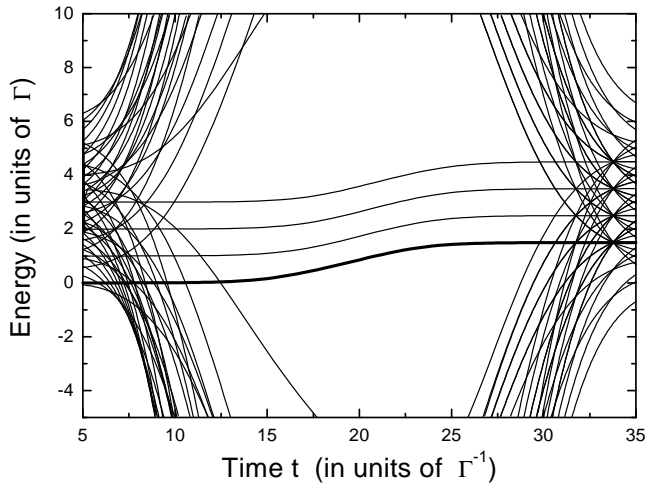,width=\hsize}}\par\vspace{5mm}
\caption{Energy eigenvalues of $H_{\text{int}}$ as a function of 
transit time for $\delta=0.5\Gamma$. The other parameters are the same
as in Fig.~\ref{Fig:EV}. The bold line represents the 
single-polarization dark state $|E_0\rangle$. While at the 
center the splitting of the dark space manifold is well separated
from all other levels, the level structure at the edges is 
rather complex. Note the degeneracy at $\Gamma t\approx34$.}
\label{Fig:DressedDelta}}
\end{figure}

For the discussion of a method to suppress diabatic transitions to
states with the wrong polarization, it is important to note the
dependence of the integral in Eq.~(\ref{LandauZener}) on the phase
factor. When $|E_0\rangle$ and $|E_1\rangle$ are degenerate, this phase
factor is obviously zero. There is, however, a simple way to lift this
degeneracy: a small detuning $\delta$ of the cavity with respect to the
pump field and the atomic transition frequency (assumed to be equal for
both polarization modes, i.e., $\delta_+=\delta_-=\delta$) will split
the dark space levels into nondegenerate eigenstates, according to
their total photon number. Therefore, these new eigenstates are exactly
the states given by Eq.~(\ref{eigenstates}). The corresponding dark
space eigenvalues form a series of energy levels separated by
approximately $2\delta$, since neighboring levels differ by two photons.
Through adiabatic passage each state acquires exactly three photons, so
that all eigenvalues undergo an increase of $3\delta$ during the atom's
transit through the cavity. This is apparent in the central part of
Fig.~\ref{Fig:DressedDelta}.

When the degeneracy of the dark space is thereby lifted, the phase
factor apppearing in Eq.~(\ref{LandauZener}) no longer vanishes, but
leads to oscillations of the function under the integral. As a
consequence, the transition probability $|c_1|^2$ will drop when
$\delta$ is increased. This simple model is in excellent agreement
with the results of a numerical integration of the full interaction 
Hamiltonian~(\ref{Hamilt}).

With the transition rate strongly suppressed at large $\delta$, a system
prepared in the $|g_{-3},0,0\rangle$ state will adiabatically follow the
$|E_{0}\rangle$ dressed state in the dark space and end up in the
$|g_{0},0,3\rangle$ state, i.e., a Fock state of the cavity field with
pure $\sigma_-$ polarization. In fact, the structure of the phase factor
in Eq.~(\ref{LandauZener}) implies a new adiabaticity condition for the
energy splitting of the dark state manifold and the time $T$ for which
the interaction is effective:
\begin{equation}
\label{newadiabcond}
|E_1-E_0| T \approx 2 |\delta| T \gg 1.
\end{equation}
This requirement must be met in addition to the usual (single-polarization)
conditions for adiabatic following \cite{Parki93,Oreg84}
\begin{equation}
\label{oldadiabcond}
\Omega_0T\gg1, \qquad 2g_0\sqrt{n_\pm+1}\; T\gg 1,
\end{equation}
which minimize the probability for diabatic transitions to ac-shifted
(nondark) dressed states. Since for typical parameters
$\delta\ll\Omega_0,g_0$, the new condition~(\ref{newadiabcond}) is more
stringent than the standard adiabatic condition~(\ref{oldadiabcond}), so
that considerably slower atoms (by a factor of $g_0/\delta$) are needed
in the two-polarization case.

%%%%%%%%%%%%%%%%%%%% Fig 6 %%%%%%%%%%%%%%%%%%%%%%%%%%%%%%%%%%%%%%%
%%%%%%%%%%%%%%%%%%%% Fig 6 %%%%%%%%%%%%%%%%%%%%%%%%%%%%%%%%%%%%%%%

\begin{figure}[tb]
\vbox{
\centerline{\psfig{file=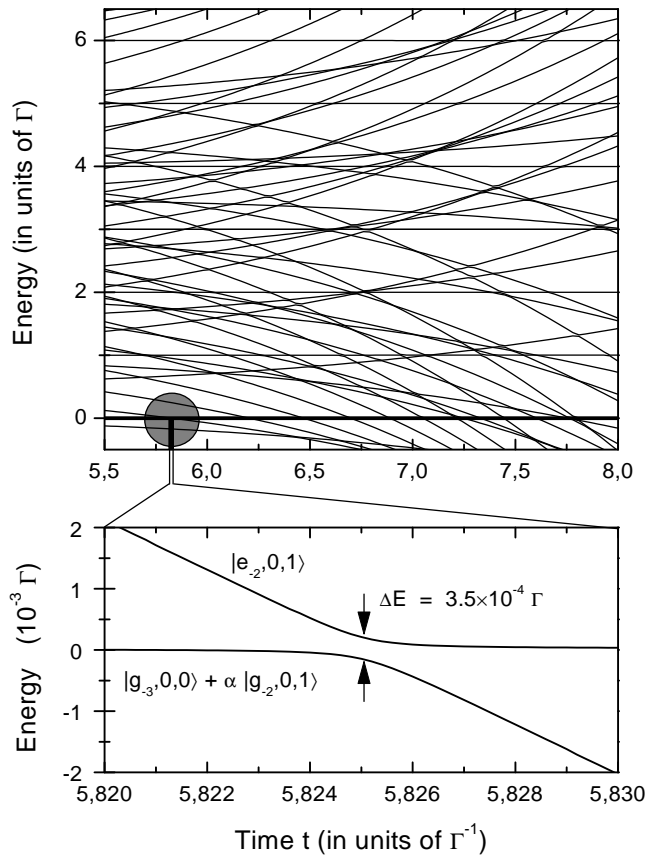,width=\hsize}}\par\vspace{5mm}
\caption{Close-up view of energy eigenvalues of $H_{\text{int}}$
at the leading edge of the cavity mode for the parameters 
of Fig.~\ref{Fig:DressedDelta}. The enlargement of the first crossing
of the lowest dark state (with state $|e_{-2},0,1\rangle$)
shows that the crossing is actually avoided on a very fine
scale.}
\label{Fig:Avoided}}
\end{figure}

In the discussion, so far it was assumed implicitly that the energies of
nondark dressed states (ac-Stark shifted on the order of $g_0$ and
$\Omega_0$) and the dark space eigenvalues (on a scale of $|\delta|$)
are well separated, as is the case in the middle of the atomic transit.
However, this assumption is not valid at the leading edge of the cavity
mode and the trailing edge of the pump field, where the ac-Stark shifts
of the energy levels become comparable to the splitting of the dark
states ($g(t),\Omega(t)\approx|\delta|$). In this regime, a complex 
structure of the eigenvalues ensues, as is apparent from the edges 
of the dressed-state diagram of Fig.~\ref{Fig:DressedDelta}. 
An enlarged view of the energy levels
for the atom entering the cavity mode (Fig.~\ref{Fig:Avoided}) shows
that the lower dark state $|E_0\rangle$, which is initially occupied,
undergoes a large number of crossings with nondark levels. Most of
these crossings are actually avoided, as illustrated in the blow up of
the first crossing in Fig.~\ref{Fig:Avoided}. Since the synthesis of
pure Fock states requires that the system follows the lowest dark state,
transitions to any intersecting states must be avoided, i.e., all
crossings must be passed {\em diabatically}. This condition puts an {\em
upper} bound on the products appearing in
Eqs.~(\ref{newadiabcond}), (\ref{oldadiabcond}), limiting them to a finite
range of values.

%%%%%%%%%%%%%%%%%%%% Fig 7 %%%%%%%%%%%%%%%%%%%%%%%%%%%%%%%%%%%%%%%
%%%%%%%%%%%%%%%%%%%% Fig 7 %%%%%%%%%%%%%%%%%%%%%%%%%%%%%%%%%%%%%%%

A quantitative evaluation of the eigenvalue spectra shows that the
effects of level crossings are different when the atom is mainly exposed
to the cavity field (in the initial phase of the atomic transit with
$g/\Omega\gg 1$) and when the interaction is primarily with the pump
field (in the final phase with $g/\Omega\ll 1$). In the former case, it
should be noted that only two-level crossings occur. The largest level
splitting is found at the first crossing, which is the one shown in the
lower part of Fig.~\ref{Fig:Avoided}. Corresponding to the small
coupling at the edge of the cavity mode, it has an energy splitting
orders of magnitude smaller than $\Gamma$ ($3.5\times10^{-4}\Gamma$ in
the example of Fig.~\ref{Fig:Avoided}). In order to pass this crossing
adiabatically, the atom would have to have a velocity $10^6$ times lower
than that taken for Fig.~\ref{Fig:EV}. Therefore, for realistic atomic
velocities the diabatic condition is easily fulfilled on entry and the
atom will pass the crossings as if the levels were not coupled. This
result is confirmed by the numerical calculations of the following
section.

%%%%%%%%%%%%%%%%%%%% Fig 8 %%%%%%%%%%%%%%%%%%%%%%%%%%%%%%%%%%%%%%%%
\begin{figure}[tb]
\vbox{
\centerline{\psfig{file=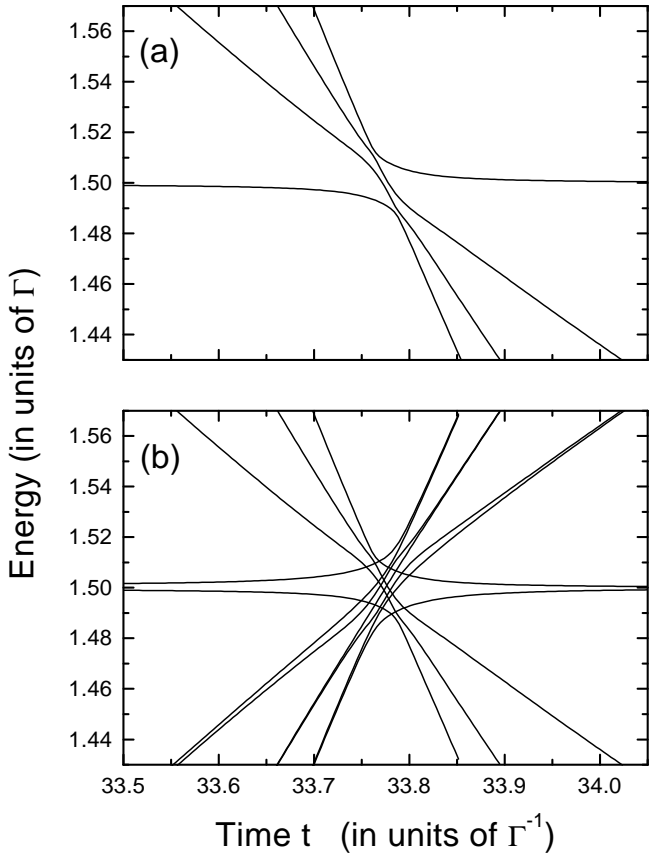,width=\hsize}}\par\vspace{5mm}
\caption{Dressed level structure at the trailing edge of the
pump pulse where $\Omega(t)\approx 4|\delta|$. A degeneracy of the
atom-pump interaction leads to a multilevel avoided crossing.
(a) Single-polarization case; (b) two-polarization case.
The parameters are the same as in Fig.~\ref{Fig:DressedDelta}.
}
\label{Fig:Degenerate}}
\end{figure}
%%%%%%%%%%%%%%%%%%%% Fig 8 %%%%%%%%%%%%%%%%%%%%%%%%%%%%%%%%%%%%%%%

A different situation occurs at the trailing edge of the pump field. At
a time $t$ such that the classical Rabi frequency has decreased to
$\Omega(t)=4|\delta|$, the ac-shifted dressed levels become degenerate
with the series of dark states (neglecting coupling to the cavity mode),
as is apparent from Fig.~\ref{Fig:DressedDelta} and shown in more detail
in Fig.~\ref{Fig:Degenerate} for the lowest dark state. The phenomenon
is found in the single- [Fig.~\ref{Fig:Degenerate}(a)] and the
two-polarization case [Fig.~\ref{Fig:Degenerate}(b)] at finite detuning.
The degeneracy is lifted due to the interaction with the cavity mode(s),
so that a multiple-level avoided crossing is created. For positive
detuning $\delta$ and a single polarization, the multiplicity is given
by the number of dressed states with positive ac-Stark shift [three
levels crossing the dark state in the example of
Fig.~\ref{Fig:Degenerate}(a)]. In order to emerge from the crossing in
the dark state, the system would have to undergo diabatic jumps across
three levels with a substantial energy splitting (on the order of
$0.02\Gamma$ for $\delta=0.5\Gamma$ in the example of
Fig.~\ref{Fig:Degenerate}). Consequently, there is a finite probability
for transitions to states outside the dark state manifold. These partly
adiabatic transitions at the trailing edge of the pump field limit the
detuning $\delta$ that may be applied to achieve adiabatic conditions at
the center of the interaction region. They also set an upper limit on
the transit time $T$ and hence on the products of
Eqs.~(\ref{newadiabcond}) and~(\ref{oldadiabcond}), as pointed out
earlier.

For two polarizations, an additional degeneracy occurs due to
contributions from states with higher photon number. In this case
dressed states with negative and positive Stark shift are involved in
the crossing [Fig.~\ref{Fig:Degenerate}(b)]. Note that a similar
multiple-level crossing does not occur when the atom enters the cavity,
as the dressed energies of the quantized cavity mode are nondegenerate
due to their dependence on the photon number. Hence the restrictions for
detuning and time of flight are more stringent during interaction with
the classical field in the final phase of the atomic transit.

In the following sections, the preceding conclusions drawn from the
eigenvalue spectrum of the Hamiltonian in Eq.~(\ref{Hamilt}) and a
simple two-level model will be put on a quantitative basis by
integrating the master equation of the system, including dissipative
processes in the treatment. A suitable method is the quantum Monte Carlo
wave-function technique. The results will be employed to describe a
strategy and calculate the efficiency for the synthesis of pure
$|0,3\rangle$ or $|3,0\rangle$ states of the cavity field or
superpositions of both. 

\section{Quantum Monte Carlo simulations}
\label{Sec:QMC}

In the previous analysis only the coherent interaction of the atom and the
two cavity modes was considered. A realistic model of the system also has to
take into account dissipation that enters due to spontaneous decay of
the atom and damping of the cavity mode. The dissipative atom-cavity
system is described by the following master equation for the density
operator $\rho(t)$ of the system:
\begin{eqnarray}
\label{master}
\frac{\partial\rho }{\partial t}&=&
-\frac{i}{\hbar}\left[H_{\text{eff}},\rho\right]
+2\kappa (a_+ \rho a_+^\dagger + a_- \rho a_-^\dagger)\nonumber\\
&& +\Gamma\sum_{\sigma = 0,\pm1} A_{\sigma} \rho A_{\sigma}^\dagger.
\end{eqnarray}
Here, $H_{\text{eff}}$ is a non-Hermitian
effective Hamiltonian including decay terms given by
\begin{equation}
\label{Heff}
H_{\text{eff}}=H_{\text{int}}-i\kappa
(a_+^\dagger a_+ + a_-^\dagger a_-)
-i{\Gamma\over2}\sum_{\sigma = 0,\pm1} A_{\sigma}^\dagger A_{\sigma}.
\end{equation}
Solutions of the master equation may be obtained by integrating
Eq.~(\ref{master}) numerically. However, the size of the Hilbert space
accessible to the system in the two-polarization case is rather large,
even if its basis is truncated at moderate photon numbers. For example,
for atomic levels with $F_g=F_e=3$ and a cutoff photon number of seven in
both cavity modes, the dimension $N$ of the Hilbert space is
$2\times(2\times3+1)\times 8\times 8=896$. The computational expense
involved in density matrix calculations, which scales as $N^4$, makes
integrating the master equation impractical. A more suitable approach is
to simulate the master equation by using Monte Carlo
wave-functions\cite{Dum92a,Molme93,Carmi93}. Here, only the Schr\"odinger
equation with the non-Hermitian Hamiltonian $H_{\text{eff}}$ has to be
integrated and the problem size scales as $N^2$. Dissipation is taken
into account by {\em quantum jumps}, which occur at random times with a
probability determined by the decay rate of the wave-function norm. The
jumps are implemented by applying so-called {\em collapse operators} to
the system wave-function and normalizing the result. Between subsequent
jumps, the system evolves according to the Schr\"odinger equation with
the non-Hermitian effective Hamiltonian $H_{\text{eff}}$. In this way,
one obtains a wave function conditioned on a particular decay history.
The system density matrix and operator expectation values are obtained
by repeating the calculation many times and averaging over the resulting
trajectories. Details of the procedure are found in
Refs.~\cite{Dum92a,Molme93,Carmi93,Dum92}.

%%%%%%%%%%%%%%%%%%%% Fig 9 %%%%%%%%%%%%%%%%%%%%%%%%%%%%%%%%%%%%%%%%
\begin{figure}[tb]
\vbox{
\centerline{\psfig{file=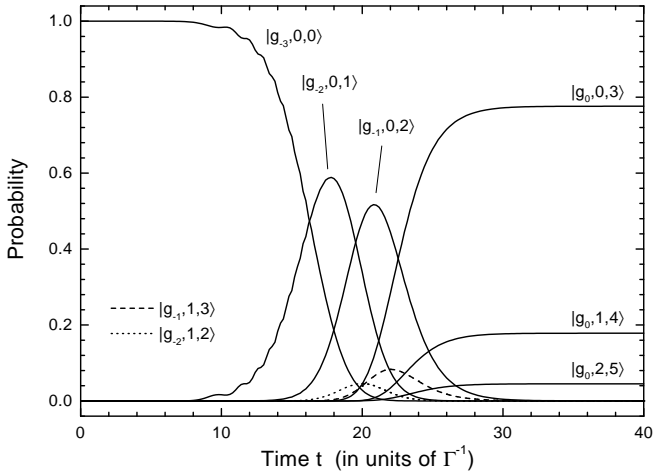,width=\hsize}}\par\vspace{5mm}
\caption{System evolution of an atom and two degenerate polarization modes,
obtained from a Monte Carlo wave-function simulation including
atomic spontaneous emission. Shown is the occupation
probability of the basis states of the system.
The amplitudes of the Gaussian pulses are $g_0=25\Gamma$ 
(at $t=17\Gamma^{-1}$) and $\Omega_0=50\Gamma$ (at $t=23\Gamma^{-1}$),
with a FWHM of $10\Gamma^{-1}$. The cavity is at resonance with
the pump field ($\delta_\pm=\omega_{c\pm}-\omega_l=0$).}
\label{Fig:MCzero}}
\end{figure}
%%%%%%%%%%%%%%%%%%%% Fig 9 %%%%%%%%%%%%%%%%%%%%%%%%%%%%%%%%%%%%%%%%

In addition to the potential for more efficient use of computational
resources, the method offers the benefit that for each of the simulated
trajectories a record of the quantum jumps that have occurred during the
system evolution is obtained. In the case of Eq.~(\ref{master}), there
are five possible collapse operators and hence five types of jumps given
by
\begin{equation}
\sqrt{2\kappa}\; a_\pm ,\qquad\sqrt{\Gamma}\; A_\sigma,\qquad\sigma = 0,\pm1,
\end{equation}
corresponding to decay of the two orthogonally polarized cavity modes
and spontaneous emission of a photon with polarization $\sigma$,
respectively. Decay of the cavity modes may be associated with
photodetection events in an individual experimental run. Note that, when
the polarization of the photons leaking from the cavity is analyzed in a
basis different from that of the intracavity field (as set by the
spatial quantization of the atom), linear combinations of the field
decay operators $a_{\pm}$ must be employed \cite{Mabuc96a}, which will
be used in Sec.~\ref{Sec:GHZdet} to analyze GHZ correlations between
cavity photons.

In a first set of simulations, only atomic decay is included, while
cavity decay is still neglected. Ideally, as long as the evolution of
the system is restricted to the dark space, the influence of spontaneous
emission on the dynamics should be weak, since the upper atomic levels
are not excited. Hence, the conclusions of Sec.~\ref{Sec:States} should
apply to the numerical results in this regime.

We start by studying the system evolution for an $F_g=3\rightarrow
F_e=3$ transition in the case of atom, cavity, and laser fields on
resonance ($\delta=0$), i.e., the case of a degenerate dark space. The
initial state is again $|g_{-3},0,0\rangle$. Figure~\ref{Fig:MCzero}
shows the system evolution during the passage of an atom through the
time-dependent cavity coupling $g(t)$ and the laser field $\Omega(t)$ in
terms of the basis state occupation. The probability maximum is shifted
in turn from the initial state to $|g_{-2},0,1\rangle$ to
$|g_{-1},0,2\rangle$, before finally reaching $|g_{0},0,3\rangle$. These
are exactly the levels comprised by the lowest dressed state
$|E_0\rangle$ in the dark space manifold of Eq.~(\ref{eigenstates}),
which should be the only ones occupied if the interaction was entirely
adiabatic. Approximately halfway through the transit, however,
contributions from levels $|g_{0},1,4\rangle$ and $|g_{0},2,5\rangle$
start to appear. These are due to the diabatic transitions to states
with higher photon number predicted from the two-level model analyzed in
Sec.~\ref{Sec:States}. At the end of the transit, the atom is found in
state $|g_{0}\rangle$ and the cavity is in a superposition of the states
$|0,3\rangle$, $|1,4\rangle$, and $|2,5\rangle$, etc., instead of the
desired pure Fock state $|0,3\rangle$, confirming the earlier result of
Eq.~(\ref{finalstate}).

%%%%%%%%%%%%%%%%%%%% Fig 10 %%%%%%%%%%%%%%%%%%%%%%%%%%%%%%%%%%%%%%%
\begin{figure}[tb]
\vbox{
\centerline{\psfig{file=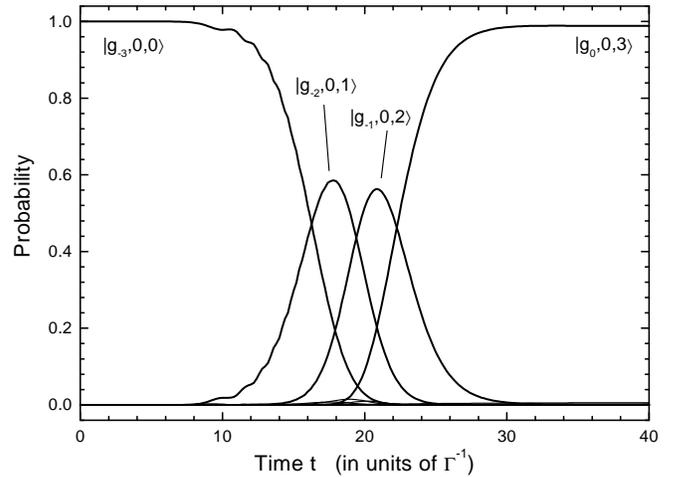,width=\hsize}}\par\vspace{5mm}
\caption{System evolution of atom and two polarization modes,
at a finite cavity detuning $\delta=0.6\Gamma$, showing 
nearly ideal adiabatic passage. 
The amplitudes and timing of the applied pulses are the
same as in Fig.~\ref{Fig:MCzero}.}
\label{Fig:MCdelta}}
\end{figure}
%%%%%%%%%%%%%%%%%%%% Fig 10 %%%%%%%%%%%%%%%%%%%%%%%%%%%%%%%%%%%%%%%

In the two-level model of Eqs.~(\ref{twolevelSE}), (\ref{LandauZener}) in
Sec.~\ref{Sec:States}, a finite cavity detuning $\delta$ was shown to
lift the degeneracy of the dark space and lead to a strong suppression
of diabatic transitions. What remains to be investigated is the question
of whether this mechanism is still effective, if the full system
dynamics including atomic decay and level crossings in the entrance and
exit phase of the interaction are taken into account. In
Fig.~\ref{Fig:MCdelta}, a Monte Carlo calculation of the system
evolution for a cavity detuning $\delta_\pm=0.6\Gamma$ is shown. The
pump field is still assumed to be on resonance with the atomic frequency
($\omega_l=\omega_a$). As can be seen from the resulting final state,
the finite cavity detuning leads to the desired strong suppression of
states with $\sigma_+$ photons and an amplitude for the
$\sigma_-$ Fock state, which is close to unity (99\%). The atom remains
in the lowest branch of the dark state manifold and is adiabatically
transferred from $|g_{-3},0,0\rangle$ to $|g_0,0,3\rangle$. Therefore,
passage through a detuned cavity is adiabatic and well suited to
generate a Fock state of one polarization of the cavity field, even for
a cavity supporting two degenerate polarization modes. By choosing other
initial states ($|g_{-m},0,0\rangle,\; m=1,2$), states with photon
number $m<F_g$ may be obtained as well. Coherent superpositions of
Zeeman ground states with $0<m\le F_g$ are also possible, leading to
superpositions of photon number states. Thus an operationally simple
modification of the work of Ref.~\cite{Parki93} suffices to make a
workable scheme for the generation of ``arbitrary" field states for a
single polarization mode.

%%%%%%%%%%%%%%%%%%%% Fig 11 %%%%%%%%%%%%%%%%%%%%%%%%%%%%%%%%%%%%%%%
\begin{figure}[tb]
\vbox{
\centerline{\psfig{file=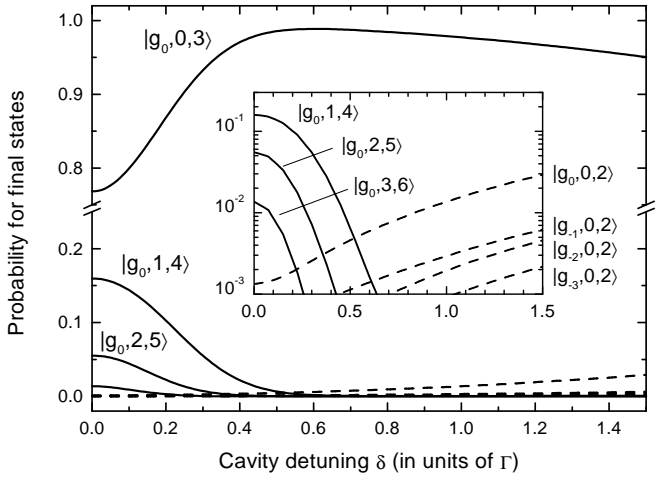,width=\hsize}}\par\vspace{5mm}
\caption{Final state probabilities as a function of cavity
detuning $\delta_+=\delta_-=\delta$, showing the transition to an adiabatic
regime for $\delta>0.5\Gamma$. The parameters correspond 
to those of Fig.~\ref{Fig:MCzero}.}
\label{Fig:detuning}}
\end{figure}
%%%%%%%%%%%%%%%%%%%% Fig 11 %%%%%%%%%%%%%%%%%%%%%%%%%%%%%%%%%%%%%%%

To study systematically the effect of cavity detuning $\delta$ (assumed
to be equal for both polarizations, i.e., $\delta_+=\delta_-=\delta$) on
the efficiency of our scheme, we have calculated the final state as a
function of $\delta$, with the result displayed in
Fig.~\ref{Fig:detuning}. As $\delta$ is increased from zero, the
probability for contributions from basis states with the ``wrong" cavity
polarization ($\sigma_+$) drops to zero, while the pure $\sigma_-$ state
becomes occupied with 99\% efficiency around $\delta\approx 0.6\Gamma$.

In contrast to the results of the two-level model, the probability for a
pure $\sigma_-$ Fock state drops again slightly when $\delta$ is
increased to values larger than $0.6\Gamma$. The reason is the dynamics
in the initial and the final phase of the atomic transit (for example, at
the multilevel crossing shown in Fig.~\ref{Fig:Degenerate}), which
becomes partly adiabatic if $\delta$ is chosen too large. As a
consequence, population is transferred to levels outside the dark state
manifold, leading not only to a finite probability for a decreased
photon number in the cavity, but also to entanglement between the atom and
the cavity field. This is illustrated in Fig.~\ref{Fig:detuning} by the
rise of contributions from levels $|g_{k},0,2\rangle$ ($k=-3,\dots,0$)
as $\delta$ is increased beyond $0.6\Gamma$. Note that this limitation
of adiabatic passage also occurs in the single-polarization case and is
not connected to the structure of the dark space.

%%%%%%%%%%%%%%%%%%%% Fig 12 %%%%%%%%%%%%%%%%%%%%%%%%%%%%%%%%%%%%%%%
\begin{figure}[tb]
\vbox{
\centerline{\psfig{file=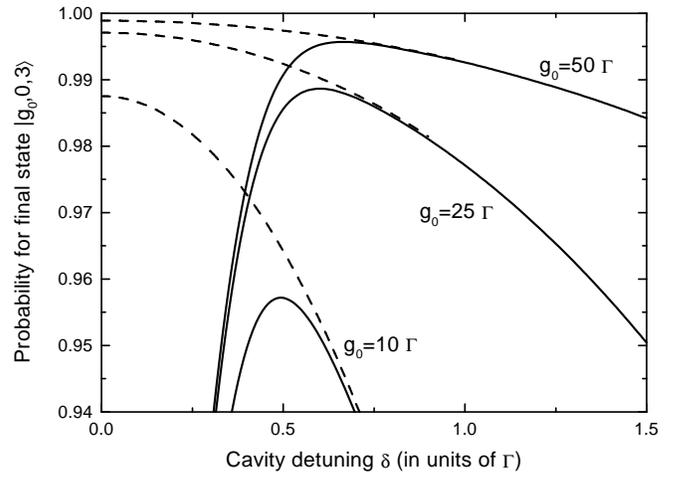,width=\hsize}}\par\vspace{5mm}
\caption{Probability for a final state $|g_0,0,3\rangle$
as a function of cavity detuning $\delta$ for three coupling
constants in the case of two-polarization adiabatic passage 
(solid lines). The single-polarization ($\sigma_-$) results 
are included as well (dotted lines). The timing of the pulses 
is the same as in Fig.~\ref{Fig:MCzero} and $\Omega_0=2g_0$.
Note that, for this figure, the full master equation was simulated 
in order to obtain an accuracy well below 1\%.}
\label{Fig:gdetune}}
\end{figure}
%%%%%%%%%%%%%%%%%%%% Fig 12 %%%%%%%%%%%%%%%%%%%%%%%%%%%%%%%%%%%%%%%

A blown-up plot of the probability for reaching the final state
$|g_0,0,3\rangle$ vs detuning $\delta$ is presented in
Fig.~\ref{Fig:gdetune}. The results for three different values of the
coupling constant $g_0$ show that the efficiency of the method
approaches unity when the coupling is increased (99.6\% at
$g_0=50\Gamma$). The optimum detuning is $\delta\approx0.6\Gamma$ and
depends on the coupling only weakly. For comparison, the
single-polarization probability for a three-photon Fock state is also
included in Fig.~\ref{Fig:gdetune}. The large deviation between the two
sets of curves for values below $\delta=0.6\Gamma$ clearly shows the
inadequacy of the single-polarization model. For large detunings, on the
other hand, the results coincide, demonstrating that in this regime the
two-polarization dynamics is well described by the adiabatic following
of a single dark state. The common asymptotic behavior of the curves is
another hint to the polarization independent mechanism populating
nondark states at the multilevel avoided crossing occurring for large
$\delta$ (Fig.~\ref{Fig:Degenerate}). Figure~\ref{Fig:gdetune} suggests
that, for an optimum value of $\delta$, the purity of the final state is
only limited by the atom-cavity coupling $g_0$ that may be achieved
experimentally, if the interaction time is held constant.

%%%%%%%%%%%%%%%%%%%% Fig 13 %%%%%%%%%%%%%%%%%%%%%%%%%%%%%%%%%%%%%%%
\begin{figure}[tb]
\vbox{
\centerline{\psfig{file=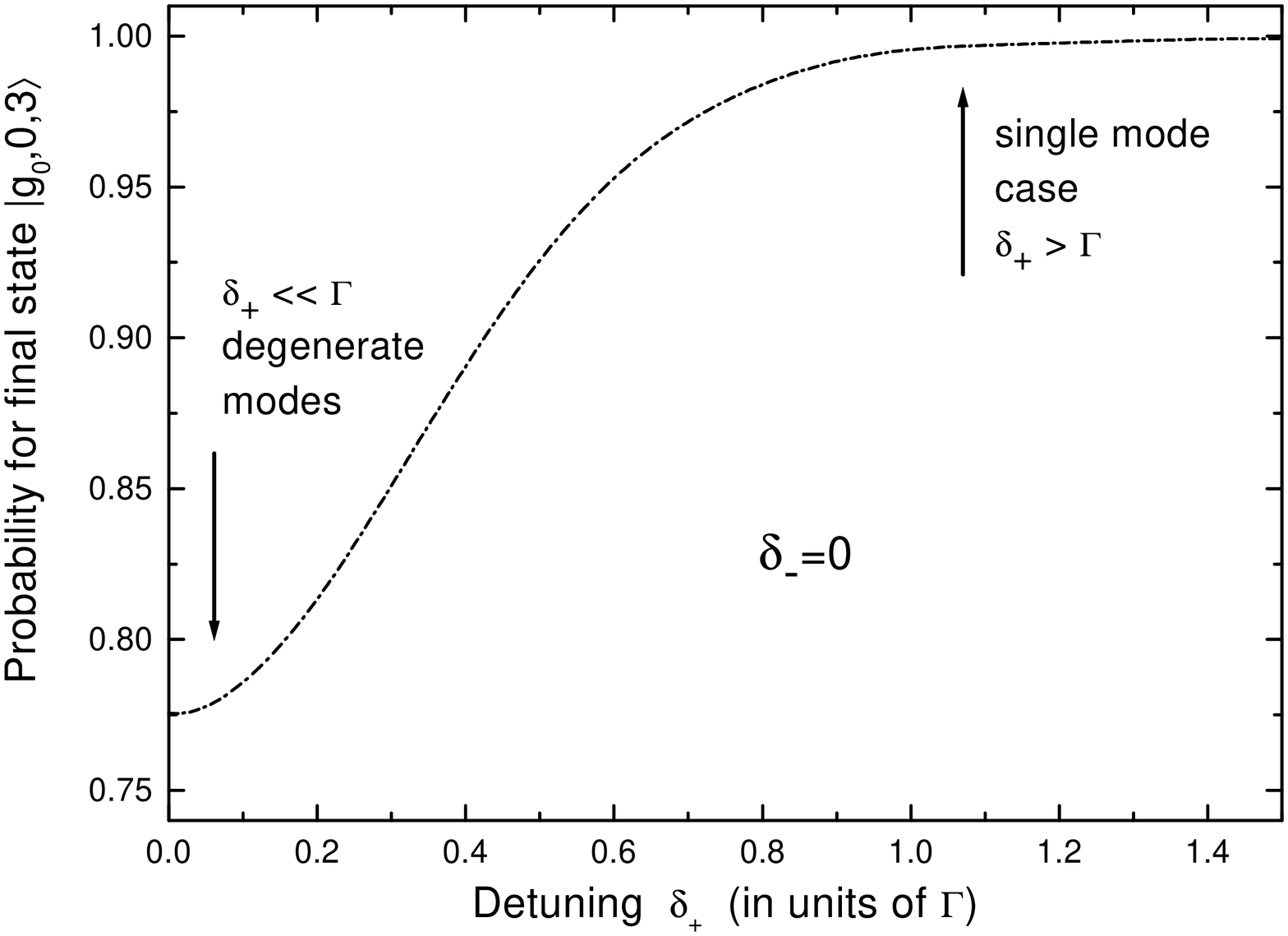,width=\hsize}}\par\vspace{5mm}
\caption{Probability for a final state $|g_0,0,3\rangle$
as a function of the detuning $\delta_+$ of the $\sigma_+$ polarized
cavity mode. The $\sigma_-$ mode is assumed to be on resonance and
a coupling $g_0=25\Gamma$ was used for both modes.
The timing of the pulses is the same as in Fig.~\ref{Fig:MCzero}.}
\label{Fig:gdetuneplus}}
\end{figure}
%%%%%%%%%%%%%%%%%%%% Fig 13 %%%%%%%%%%%%%%%%%%%%%%%%%%%%%%%%%%%%%%%

In the calculations presented so far, the detunings of the two
polarization modes of the cavity were always assumed to be equal
($\delta_+=\delta_-$). In order to probe the sensitivity of the system
to a finite mode splitting, which may occur due to cavity imperfections,
we have varied the detuning $\delta_+$ of the $\sigma_+$-polarized mode
while keeping the $\sigma_-$-polarized mode at resonance with the atom
($\delta_-=0$). Again, for the initial state $|g_{-3},0,0\rangle$,
Fig.~\ref{Fig:gdetuneplus} shows the probability to reach a final
state $|g_0,0,3\rangle$ as a function of $\delta_+$. It exhibits a
gradual transition from the diabatic losses in the degenerate
two-polarization case to the perfect adiabatic passage obtained for a
single ($\sigma_-$) polarization, as $\delta_+$ is increased from zero.
The single mode description applies
when the $\sigma_+$ mode is far enough detuned from resonance for its
influence on the system dynamics to be neglected, which is the case for
detunings $\delta_+\gtrsim\Gamma$. Figure~\ref{Fig:gdetuneplus} suggests
that, for the model of degenerate modes to be valid, the splitting
should be small compared with the natural linewidth $\Gamma$. As long as
$\delta_+<0.1\Gamma$, the final state probability deviates from the
degenerate case by less than a percent. Note that for realistic cavities
a well-defined splitting in the $\sigma_ \pm$ basis may be difficult to
achieve, because of a lack of control over the polarization basis in
which birefringence of the mirrors occurs.

At fixed coupling $g_0$, the efficiency of the state synthesis scheme
may be increased by extending the time $T$ during which the atom
interacts with the cavity mode and the pump field, i.e., using slower
atoms for the adiabatic passage. In Sec.~\ref{Sec:States} it was found
that adiabatic transitions to nondark states place an upper limit on
$T$. There is, however, another restriction for the interaction time,
which is set by the decay time of the cavity modes. This will be
demonstrated in the following section by including cavity decay in the
master equation.

\section{Preparation of a Greenberger-Horne-Zeilinger state}
\label{Sec:GHZprep}

Adiabatic passage with a detuned cavity and an atom prepared in the
$|g_{-3}\rangle$ level was shown in Sec.~\ref{Sec:QMC} to synthesize
Fock states of $\sigma_-$ polarization and effectively suppress excitation of
the $\sigma_+$-polarized cavity mode. An important property of the proposed
scheme is that it is entirely symmetric under a change of sign in the magnetic
quantum number and a simultaneous exchange of $\sigma_-$  and
$\sigma_+$ polarization, as is apparent from the master equation~(\ref{master})
and the relevant Clebsch-Gordan coefficients. If, for example, the system is
prepared in the state $|g_{+3},0,0\rangle$, a Fock state with three
$\sigma_+$ photons in the cavity ($|3,0\rangle$) will be generated. Another
notable feature of the two polarization adiabatic passage for an $F\rightarrow
F$ transition is that at the end of the interaction the atom is always in the
$m_g=0$ Zeeman ground state ($|g_0\rangle$), independent of the initially
occupied Zeeman substates. Therefore, the atomic part of the wave function is
always disentangled from the state of the cavity field and can be disregarded
without introducing projection noise to the measurement of the cavity field.
The method is not restricted to single Zeeman levels as initial atomic states.
The atom may be prepared in a coherent superposition of Zeeman ground states,
leading to superposition states of the cavity field as described for the 
single-polarization 
model in Refs.~\cite{Parki93,Parki95}. A particularly intriguing
possibility in the two-polarization case is to use coherent superpositions of
Zeeman sublevels with different sign of the magnetic quantum number $m$. In
this case, quantum superpositions of photon states with different polarization
are synthesized, i.e., the adiabatic passage maps the Zeeman coherence of the
atomic state to polarization entanglement of the two cavity modes. 

%%%%%%%%%%%%%%%%%%%% Fig 14 %%%%%%%%%%%%%%%%%%%%%%%%%%%%%%%%%%%%%%%
\begin{figure}[tb]
\vbox{
\centerline{\psfig{file=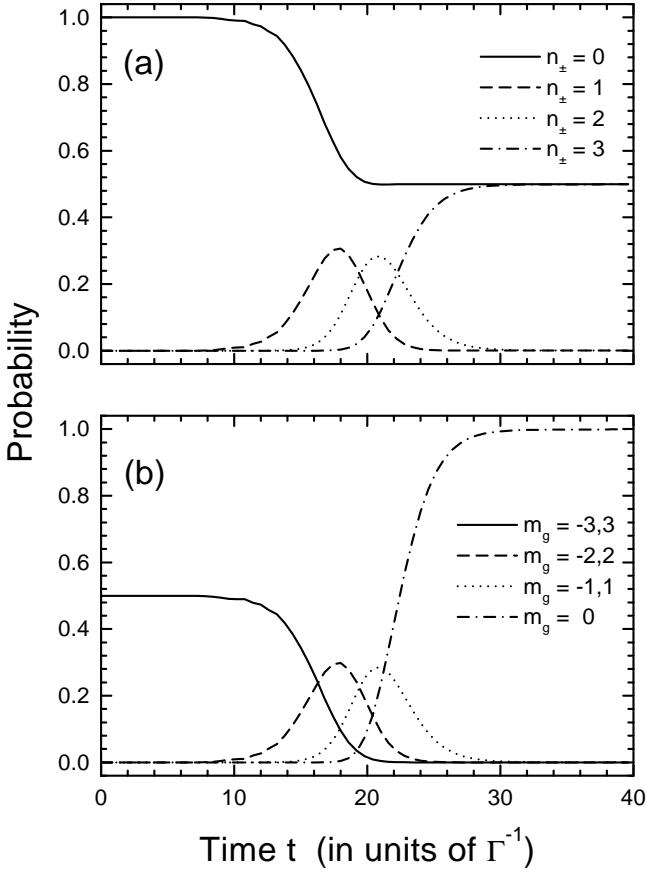,width=\hsize}}\par\vspace{5mm}
\caption{Monte Carlo simulation of the preparation of a GHZ-type state
in the absence of cavity decay. The amplitudes of the cavity and 
pump field are $g_0=25\Gamma$ (at $t=17\Gamma^{-1}$) and 
$\Omega_0=50\Gamma$ (at $t=23\Gamma^{-1}$) with a FWHM of $10\Gamma^{-1}$,
the cavity detuning is $\delta=0.6\Gamma$, and the initial preparation
of the atom is a coherent superposition of $m_g=3$ and $m_g=-3$.
(a) Occupation of the cavity modes (the distributions are identical 
for the $\sigma_+$  and the $\sigma_-$ mode);
(b) atomic ground-state populations (the curves for $+m_g$ and $-m_g$
coincide). The final state is a 99\% pure state $|\text{GHZ}\rangle$.}
\label{Fig:GHZ}}
\end{figure}
%%%%%%%%%%%%%%%%%%%% Fig 14 %%%%%%%%%%%%%%%%%%%%%%%%%%%%%%%%%%%%%%%

An important example of such field states are the number states of $n$ photons
maximally entangled in polarization, which are obtained for a superposition
state of $n$ photons in the $\sigma_+$ mode and $n$ photons in the
$\sigma_-$ mode. With the notation previously introduced for a two-mode cavity
field, these states may be written as
\begin{equation}
\label{GHZ}
|\Psi_n\rangle\equiv {1\over \sqrt{2}} \left(|n,0\rangle+|0,n\rangle\right).
\end{equation}
For the state $|\Psi _{2}\rangle$, there is a close correspondence with the
correlated two-particle states used to test Bell's inequalities
\cite{Bell65,Claus69} obtained from local hidden-variable theories. Numerous
experiments have shown these inequalities to be violated \cite{EPR}. Yet even
more interesting properties emerge from the entangled state $|\Psi_3\rangle$ of
three particles, corresponding to a so-called {\em GHZ state}, as was introduced
by Greenberger, Horne, and Zeilinger \cite{GHZ89}. Its significance is based on
the fact that, to test theories of local realism against quantum mechanics, a
single set of observations of this state is sufficient ``all or nothing"). This
is in contrast to the statistical nature of the violation of Bell's inequalities
(for $n=2$).

%%%%%%%%%%%%%%%%%%%% Fig 15 %%%%%%%%%%%%%%%%%%%%%%%%%%%%%%%%%%%%%%%
\begin{figure}[tb]
\vbox{
\centerline{\psfig{file=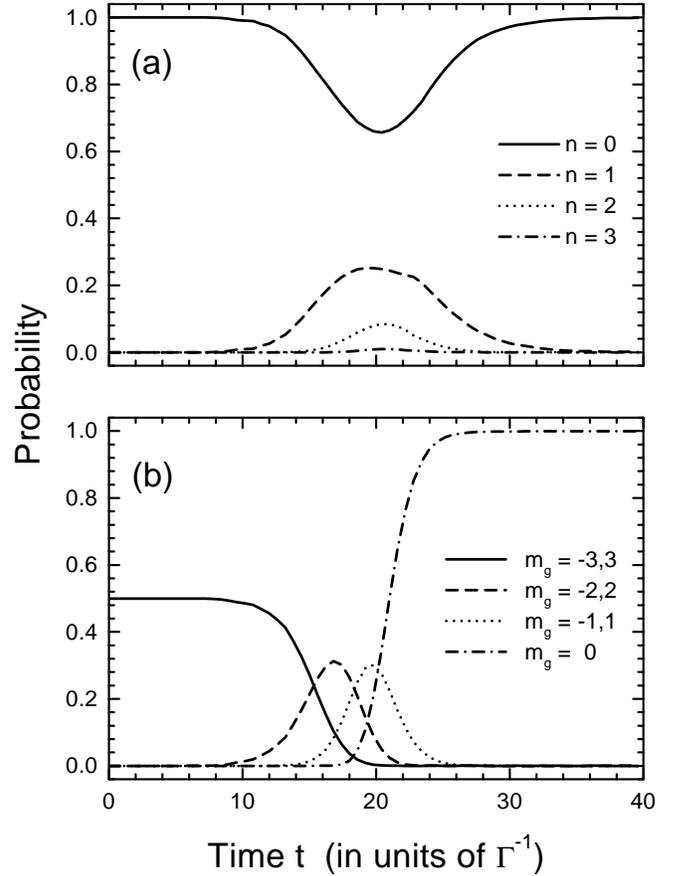,width=\hsize}}\par\vspace{5mm}
\caption{Monte Carlo simulation of adiabatic passage for the
same parameters as in Fig.~\ref{Fig:GHZ}, but with a cavity
decay rate $\kappa=0.2\Gamma$.
(a) Occupation of the cavity modes;
(b) atomic ground-state populations.}
\label{Fig:GHZdecay}}
\end{figure}
%%%%%%%%%%%%%%%%%%%% Fig 15 %%%%%%%%%%%%%%%%%%%%%%%%%%%%%%%%%%%%%%%

There is a notable difference between the states $(|n,0\rangle$ $+$ $|0,n\rangle )$
considered here and the original Bell-type states of the form
$(|+_{1},+_{2},\dots ,+_{n}\rangle + |-_{1},-_{2},\dots ,-_{n}\rangle )$. While
in the former case all photons of a given polarization occupy a single mode, in
the latter case they are distributed over n distinct (i.e., distinguishable)
modes ($1, 2,\dots, n$). However, as will be discussed below, we restrict our
attention to the subensemble of photoelectric detection events that produce
one and only one detection event (``click") at each of $n$ sets of detectors.
Hence, from the perspective of a post diction, the state $|\Psi _{n}\rangle $
for the subensemble of distinct ``temporal" modes generates correlations
identical to those of the Bell-type states. More generally, by tailoring the
interaction profiles $g(t)$ and $\Omega(t)$ of cavity and pump field, an
improved timing of cavity decay events may be achieved, so that in fact
successive quanta $1, 2,\dots, n$ would emerge from the cavity in well defined
intervals and would be routed in a predefined fashion to respective detectors,
making the requirement of separate detection events even less restrictive.

In the following, we give a detailed description of how the two-polarization
adiabatic passage scheme may be extended to generate a GHZ-type state
$|\Psi_3\rangle$ (denoted as $|\text{GHZ}\rangle$) with high efficiency.
Towards this end the $F_g=3\rightarrow F_e=3$ transition used for the
generation of the single-polarization Fock state will be employed again. Note
that states $|\Psi_n\rangle$ with $n>3$ may be synthesized as well, if an
atomic transition with a high enough angular momentum were to be used. For the
$F_g=4\rightarrow F_e=4$ hyperfine transition in cesium, for example, the state
$|\Psi_4\rangle$ may be generated in exactly the same way as $|\Psi_3\rangle$.

The starting point for the generation of a GHZ-type state is to create a
coherent superposition of the two outer Zeeman ground states ($m_g=\pm3$). This
task may be achieved by using suitably timed microwave \cite{Carte95} or Raman
pulses \cite{Law98}. Ideal adiabatic passage of an atom through the cavity
modes with coupling $g(t)$ and the pump field $\Omega(t)$, either by physical
motion or by suitable modulation of the coupling for a stationary atom in the
cavity, would then generate the following mapping of ground-state Zeeman
coherences to the cavity field:
\begin{eqnarray}
\left(|g_{-3}\rangle + |g_{+3}\rangle \right) \otimes |0,0\rangle
&\Longrightarrow &
|g_{0}\rangle \otimes  \left( |0,3\rangle + |3,0\rangle \right)
\nonumber\\
&\sim& |g_{0}\rangle \otimes |\text{GHZ}\rangle
\end{eqnarray}
and would thus leave the cavity in a maximally entangled state of the
GHZ-type. The transitions relevant for this synthesis scheme are
the ones shown in Fig.~\ref{Fig:Csscheme}(b).

We have tested the efficiency of such a GHZ-state preparation procedure
numerically by simulating the relevant master equation using Monte Carlo
wave-functions, again neglecting cavity decay in the first instance but with
atomic decay included. The same parameters as in Secs.~\ref{Sec:States} and
\ref{Sec:QMC} are used with a detuning $\delta=0.6\Gamma$ to achieve optimum
purity of the final state. The resulting averaged time evolution is shown in
Fig.~\ref{Fig:GHZ}. In Fig.~\ref{Fig:GHZ}(a) the occupation of the cavity mode states is
plotted, while Fig.~\ref{Fig:GHZ}(b) shows the corresponding atomic ground-state
populations. As expected due to the symmetry of the system, the time-dependent
photon distribution of the two cavity modes is identical. The same is true for
the occupation of Zeeman ground-state levels with the same modulus $|m_g|$. At
the end of the adiabatic passage, there is almost perfect polarization
entanglement with a 99\% probability for the cavity to be found in the GHZ
state $\left( |0,3\rangle + |3,0\rangle \right)$ and the atom in state
$|g_{0}\rangle$. In correspondence with the discussion of
Sect.~\ref{Sec:States}, contributions of higher terms in the dark space
manifold, which are again of the form $\left(|k,3+k\rangle +
|3+k,k\rangle\right)$, $k=1,2,\dots$, are strongly suppressed. As mentioned
above, the efficiency could be further increased by using an even stronger
coupling $g_0$ or slower atoms. 

Beyond these encouraging results, a realistic simulation of our experiment has
to take into account dissipation of the cavity field, which has been neglected
so far. In recent experiments cavity decay rates $\kappa/2\pi$ ranging from
$0.6$ to $35$ MHz have been achieved \cite{ICAP94,Turch95b,Mabuc96,Hood98}. As
a point of reference, here we take initially $\kappa/2\pi=1$ MHz, resulting in
$\kappa/\Gamma=0.2$, although we will subsequently investigate the dependence
of the state generation on $\kappa$. The cavity decay time $1/\kappa$ is then
comparable to the adiabatic passage time scale of maximum efficiency for state
generation [cf. Fig.~\ref{Fig:EV}(a) for the time dependence of $g(t)$
and $\Omega(t)$], so that cavity decay will play an important role in the
present scheme and must be considered in the simulation.

We have repeated the Monto Carlo wave-function simulation of the GHZ state
generation of Fig.~\ref{Fig:GHZ} using the full master equation~(\ref{master}),
including the cavity decay terms with $\kappa = 0.2\Gamma$. In
Fig.~\ref{Fig:GHZdecay}, the resulting average time evolution of the cavity
mode occupations and the atomic ground-state populations is displayed. As
expected, cavity decay strongly affects the dynamics of the field modes. During
the adiabatic passage, the probability for the excitation of photon states with
$n>0$ is substantially reduced compared to the case without cavity decay. The
probability for one photon in a cavity mode never exceeds 25\%. When the atom
leaves the cavity, both modes of the cavity field have already decayed to the
$n=0$ ground state at the rate $\kappa^{-1}$ [Fig.~\ref{Fig:GHZdecay}(a)].
Obviously, there is no time when the cavity modes approach a pure GHZ-type
state. This result is in correspondence with previous studies of the generation
of quantum superpositions in a lossy single-polarization cavity
\cite{Parki93,Parki95}.

Perhaps surprisingly, Fig.~\ref{Fig:GHZdecay}(b) shows a completely different
behavior for the atomic population. The occupation of the Zeeman states is
almost unaffected by the presence of cavity decay. The evolution proceeds
essentially undisturbed in close correspondence to the perfect cavity case
shown in Fig.~\ref{Fig:GHZ}(b). The fact that cavity dissipation does not
affect the atomic dynamics during the adiabatic passage was also observed in
the single-polarization calculations of Ref.~\cite{Parki95}. It implies that
atomic coherence is transferred without ``damage" as the cavity field undergoes
a quantum jump to a lower dark state manifold \cite{privateParki}. The
preservation of atomic coherence during the adiabatic passage even for a lossy
cavity is an essential requirement for the detection of GHZ correlations
between the photons leaking from the cavity, as will be explained in the
following section.

\section{Detection of GHZ correlations}
\label{Sec:GHZdet}

In principle, there are two different approaches to detecting nonclassical
states of the cavity field. First, the field inside the resonator may be probed
directly. This may be achieved, for example, by sending a second atom through
the cavity and reversing the adiabatic passage scheme to map the photon number
distribution of the cavity field to coherences in the Zeeman substructure
\cite{Parki93,Parki95,Walse96}. Measurements of the atomic substates would then
provide information on the state of both field modes (in principle the
``complete" state \cite{Walse96}). The required timing of the pump and
probe atoms would certainly be difficult to achieve for an atomic beam (with
Poissonian fluctuations). However, in the optical domain, the most severe
obstacle to this approach is cavity decay, since, as shown above, dissipation
of the cavity field prevents the modes from ever reaching a pure GHZ state, so
that any attempt to observe a signature of that state would fail.

This state of affairs suggests observing the photons leaking from the cavity
mirrors during decay of the field as the method of choice for detecting
GHZ correlations, especially since it affords the opportunity for tests of
local realism with the ``flying" photons that emerge from the cavity.
Correlations between these emitted photons could readily be measured by
coincidence detection, with these measurements performed in spatially separated
regions as required by the locality assumption.

Before presenting a more detailed proposal for an experimental setup to probe
the decay photons, we will first discuss the expected correlations between
photons emerging from a lossy cavity. For simplicity we start by assuming that
the two cavity modes already are in the state $|\text{GHZ}\rangle$ given by
Eq.~(\ref{GHZ}) for $n=3$. Afterwards we will examine how the correlations are
affected if dissipation is present during the state synthesis.

Greenberger, Horne, and Zeilinger have developed their argument for three
arbitrary particles in a maximally entangled state. Therefore, their analysis
can also be applied to three photons in the state $|0,3\rangle+|3,0\rangle$. To
obtain an operator with two eigenvalues $\pm1$, as considered by GHZ, we define
an observable equivalent to the spin of an atomic system. It is easy to verify
that the following operators satisfy the spin commutation relations
\begin{equation}
L_+ \equiv a_+^\dagger a_-, \quad
L_- \equiv a_-^\dagger a_+, \quad
L_z \equiv {1\over2} ( a_+^\dagger a_+ - a_-^\dagger a_- ).
\end{equation}
The operators $L_\pm$ perform a change of polarization on a cavity
photon, which is analogous to a spin flip. The observable 
corresponding to the particle analyzer in the setup proposed
by GHZ is then given by the superposition 
of $L_+$ and $L_-$ with a variable relative phase shift $\phi$, namely
\begin{equation}
\label{Ldef}
L(\phi) \equiv e^{i\phi/2}\; L_+ + e^{-i\phi/2}\; L_-,
\end{equation}
with the required eigenvalues $\pm1$.
To arrive at a measurement procedure for $L(\phi)$ in the case of
photons emitted from the cavity, we first rewrite 
\begin{equation}
\label{GHZobserv}
L(\phi) = a_x^\dagger(\phi) a_x(\phi) - a_y^\dagger(\phi) a_y(\phi).
\end{equation}
The operators $a_x(\phi)$ and $a_y(\phi)$ are annihilation operators
of linearly polarized photons in the $x$ and $y$ directions of
a coordinate system rotated by an angle $\phi/4$. They are related to 
the cavity mode operators by
\begin{eqnarray}
\label{axay}
a_x(\phi) &=& {1\over\sqrt{2}}
 (e^{-i\phi/4}\; a_+ + e^{i\phi/4}\; a_-),\nonumber \\
a_y(\phi) &=& {i\over\sqrt{2}}
 (e^{-i\phi/4}\; a_+ - e^{i\phi/4}\, a_-).
\end{eqnarray}
Therefore, according to Eq.~(\ref{GHZobserv}), $L(\phi)$ can be
determined by measuring the difference in the occupation number 
of two orthogonal linear polarization modes in a rotated frame.

To be consistent with the GHZ gedanken experiment, we must make the additional
assumption that each of the three photons from the state $|\text{GHZ}\rangle$
is delivered to a separate analyzer. In this case there are only two possible
outcomes of a measurement of $L(\phi)$: a value of $+1$ is obtained if an
$x$-polarized photon is detected, a value of $-1$ for a $y$-polarized photon. 

%%%%%%%%%%%%%%%%%%%% Fig 16 %%%%%%%%%%%%%%%%%%%%%%%%%%%%%%%%%%%%%%%
\begin{figure}[tb]
\vbox{
\centerline{\psfig{file=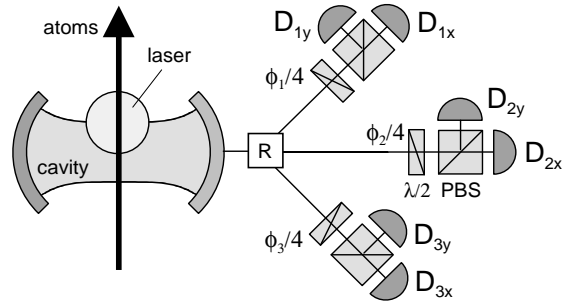,width=0.85\hsize}}\par\vspace{5mm}
\caption{Idealized experimental setup for the detection of 
GHZ correlations between photons leaking from the output port
of the cavity. Each of the three decay photons is sent by router 
R to a particular analyzer, consisting of a half-wave plate, 
a polarizing beamsplitter PBS, and a pair of detectors 
{it D}$_{\text{\bf\sf i}\xi}$ counting photons of linear 
polarization $\xi$ ($\xi$=$x$,$y$) in the basis determined by the 
wave plate.} 
\label{Fig:GHZsetupsimple}}
\end{figure}
%%%%%%%%%%%%%%%%%%%% Fig 16 %%%%%%%%%%%%%%%%%%%%%%%%%%%%%%%%%%%%%%%

Figure~\ref{Fig:GHZsetupsimple} shows an idealized experimental setup for the
detection of GHZ correlations. We assume well-defined generation times for each
of the three photons, as might be achieved in principle by a succession of
suitable shifts applied to the atomic levels in concert with a somewhat revised
time history for $g(t)$ and $\Omega (t)$ during the adiabatic passage. Each
photon emitted from the cavity is then directed to one of three analyzers. In
front of each analyzer $(k=1,2,3)$ there is a half-wave plate rotating the
respective linear polarization basis by angle $\phi_k/4$. The polarization
[$x(\phi_k)$ or $y(\phi_k)$] is measured with a linear polarizing beamsplitter
(PBS) and photon counters at the two output ports. The analyzers are assumed to
be positioned far from the cavity and from each other, so that the individual
measurements may be regarded as causally separated.

To arrive at the correlations discussed by Greenberger, Horne, and Zeilinger,
the product of the three detector results is considered. It is straightforward
to calculate its expectation value if the photons are in the GHZ state
\begin{equation}
\label{GHZop}
\langle \text{GHZ}| L_1(\phi_1) L_2(\phi_2) L_3(\phi_3) |\text{GHZ}\rangle 
= \cos\left({\phi_1+\phi_2+\phi_3}\right),
\end{equation}
where again the subscript identifies the analyzer.

The most interesting situation occurs when the detector angles are set such
that $(\phi_1+\phi_2+\phi_3)$ is an integer multiple of $\pi$. In this case the
average value is either $+1$ or $-1$, which coincides with the maximum and
minimum possible value, respectively, since any individual product of detector
outcomes is either $+1$ or $-1$. This implies that each individual measurement
of triple correlations yields the same result. In this case the photon triplets
show perfect correlations in a sense that knowing the results of two detectors
allows one to predict the outcome of the third measurement with certainty.

In Ref.~\cite{GHZ89} it was shown that the triple correlations of particles
prepared in a GHZ state violate theories based on the assumption of local
realism in a rather striking way. This may be demonstrated by considering the
correlations for the following choice of detector angles:
\begin{mathletters}
\label{violation}
\begin{eqnarray}
\left\langle\text{GHZ}\right| %
       & L_1({\pi\over 2}) L_2({\pi\over 2}) L_3(0)%
       & \left|\text{GHZ}\right\rangle = -1,\\
\left\langle \text{GHZ} \right| % 
       & L_1({\pi\over 2}) L_2(0) L_3({\pi\over 2})%
       & \left|\text{GHZ}\right\rangle = -1,\\
\left\langle \text{GHZ} \right| %
       & L_1(0) L_2({\pi\over 2}) L_3({\pi\over 2}) %
       & \left|\text{GHZ}\right\rangle = -1,\\
\left\langle \text{GHZ} \right| %
       & L_1(0) L_2(0) L_3(0)                      %
       & \left|\text{GHZ}\right\rangle = +1.
\end{eqnarray}
\end{mathletters}%

If the result of each detector was independent of the angles $\phi_k$ of the
other detectors (as required by the locality assumption for widely separated
detectors), the expectation values of Eqs.~(\ref{violation}) would factor into
the product of three functions $A_k(\phi_k)$ that can take values $\pm1$:
\begin{eqnarray}
&\langle \text{GHZ}|  L_1(\phi_1) L_2(\phi_2) L_3(\phi_3) |\text{GHZ}\rangle &
\nonumber\\
&= A_1(\phi_1)A_2(\phi_2)A_3(\phi_3)&
\end{eqnarray}
Multiplying the four left-hand sides of the set of Eqs.(\ref{violation}), a
product expression is obtained in which every factor $A_k(\phi)\; (k=1\dots3;
\phi=0,\pi/2)$ appears exactly twice. Therefore, using $A_k^2=1$), a result of
$1$ is expected for the overall left hand side product, if local realism holds.
On the other hand, the right-hand sides of Eqs.~(\ref{violation}), which were
obtained quantum mechanically, multiply to give $-1$, clearly contradicting
local realism.

We now return to the situation encountered in our experiment for realistic
parameters of coupling, transit time, and cavity decay time. It has been shown
in Sect.~\ref{Sec:GHZprep} that, even for the best Fabry-P{\'e}rot cavities available
\cite{Rempe92} (excluding for the moment the whispering gallery modes of quartz
microspheres \cite{Mabuc94b,Gorod95,Collot93}), decay may occur well before the
adiabatic passage is terminated so that the intracavity field never reaches the
state $|\text{GHZ}\rangle$. Further, since there is not a well-defined time
history for the times of the photon emissions, the operator expectation
value~(\ref{GHZop}) is not necessarily a relevant quantity. However,
correlations between the photons emitted during adiabatic passage may still be
probed. The important question then becomes whether or not they show a similar
behavior as that described by Eq.~(\ref{GHZop}).

The Monte Carlo wave-function method offers a convenient way to evaluate
correlations between photons emitted from the cavity. For each simulated
trajectory a classical record of the quantum jumps, which have occured during
the adiabatic passage, is kept that contains information on the polarization
of the decay photon corresponding to the jump. Averaging over many runs,
photon-counting distributions and correlation functions for the detectors
involved are readily calculated. To distinguish between the six possible
detection channels (two for each of the three analyzers), the collapse
operators for the two cavity modes, which appear in Eq.~(\ref{master}), must be
replaced by a new set of annihilation and creation operators. As each decay of
the cavity field can be associated with a click in a certain detector, six new
collapse operators are needed. They involve the operators defined in
Eq.~(\ref{axay}) in the rotated linear polarization bases:
\begin{eqnarray}
\sqrt{2\kappa} a_x(\phi_k)&=& %
      \sqrt{\kappa}( a_+ e^{-i\phi_k/4} + a_- e^{i\phi_k/4} );
\nonumber\\
\sqrt{2\kappa} a_y(\phi_k)&=& %
      i\sqrt{\kappa}( a_+ e^{-i\phi_k}/4 - a_- e^{i\phi_k/4} );
\label{linearbasis}
\\ && \qquad k=1,2,3.  \nonumber 
\end{eqnarray}
The new master equation to simulate is
\begin{eqnarray}
\label{newmaster}
\frac{\partial\rho}{\partial t}&=&
-\frac{i}{\hbar}\left[H_{\text{eff}},\rho\right] 
+\Gamma\sum_{\sigma = 0,\pm1} A_{\sigma} \rho A_{\sigma}^\dagger
\nonumber \\
&+& {2\kappa\over 3} \sum_{k=1,2,3} \left(a_{x}(\phi_k) 
\rho a_{x}^\dagger(\phi_k) + a_{y}(\phi_k) \rho a_{y}^\dagger(\phi_k)\right).
\end{eqnarray}
In the original GHZ proposal, three apertures are used to select the subset of
particles emitted in the direction of the three analyzers. In our case all
fields are emitted into the same spatial mode, so that one has to rely on
beamsplitters to produce three outgoing beams. In such a scheme one cannot
guarantee that photons are actually registered by separate analyzers, as
required by the GHZ argument. This was taken into account in the simulations by
discarding trajectories with a single detector being ``hit" more than once. This
way we could also discriminate against trajectories, in which more than three
photons were detected (due to occasional diabatic transitions within the dark
space manifold).

%%%%%%%%%%%%%%%%%%%% Fig 17 %%%%%%%%%%%%%%%%%%%%%%%%%%%%%%%%%%%%%%%
\begin{figure}[tb]
\vbox{
\centerline{\psfig{file=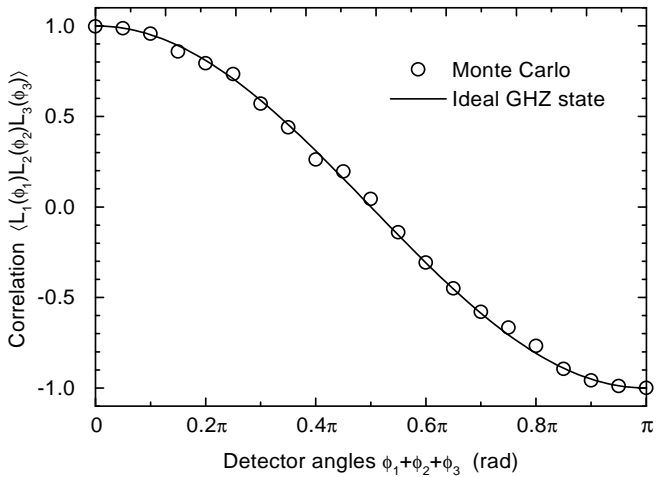,width=\hsize}}\par\vspace{5mm}
\caption{Average triple correlations of photons decaying from
the cavity, obtained from the quantum jump record of a Monte 
Carlo wave-function simulation. The adiabatic passage parameters 
are chosen as in Fig.~\ref{Fig:GHZ}, but with a finite cavity
decay rate $\kappa=0.1\Gamma$.
All three analyzers are set at the same angle ($\phi_1=\phi_2=\phi_3$),
but equivalent results are obtained for different orientations.
The solid line represents the correlations expected for an ideal
GHZ state of the cavity field in the absence of decay.}
\label{Fig:corr}}
\end{figure}
%%%%%%%%%%%%%%%%%%%% Fig 17 %%%%%%%%%%%%%%%%%%%%%%%%%%%%%%%%%%%%%%%

A comparison between the triple correlation function $\langle L_1(\phi) L_2(\phi)
L_3(\phi)\rangle$, obtained by evaluating the quantum jump record of Monte
Carlo simulations (with a decay rate $\kappa=0.1\Gamma$), and the result
calculated for an ideal GHZ state [Eq.~(\ref{GHZop})], which would have been
generated in the absence of cavity decay, is shown in Fig.~\ref{Fig:corr}.
Equal angles $\phi$ were chosen for all three detectors. Within the accuracy of
the Monte Carlo simulation results, the quantum jump correlations precisely
follow the cosine function predicted analytically for a stationary GHZ state.
In particular, perfect correlations are found for $\sum_k\phi_k=0$ and
$\sum_k\phi_k=\pi$. The Monte Carlo wave-function simulations were repeated for
different detector orientations. Whenever the sum over the angles was an
integer multiple of $\pi$, every trajectory with exactly three photons detected
showed the correlations expected from a GHZ state, even when the field started
leaking from the cavity well before the last photon was deposited to the cavity
modes by the atom.

It is remarkable that GHZ-type correlations are present between the decay
photons even when the average occupation of the cavity mode is far below one.
The observed correlations must be attributed to the dynamics of the adiabatic
passage and cannot be the property of a certain quantum state of the cavity
field [such as the case in Eq.~(\ref{GHZ})]. Preservation of coherence between
subsequent photon decays is clearly required and is provided by the Zeeman
ground states, which are not affected by decay of the cavity field (see the
time evolution of the atomic level population in Fig.~\ref{Fig:GHZdecay} in
Sec.~\ref{Sec:GHZprep}). The fact that entanglement of the ``flying" photons is
generated by the system dynamics distinguishes our method from most other
proposals for the generation of GHZ correlated particles. Another notable
feature of our scheme is that cavity dissipation does not lead to a loss of
coherence, but rather is utilized for the birth of these flying ``fields" 
with GHZ correlations into the external environment.

%%%%%%%%%%%%%%%%%%%% Fig 18 %%%%%%%%%%%%%%%%%%%%%%%%%%%%%%%%%%%%%%%
\begin{figure}[tb]
\vbox{
\centerline{\psfig{file=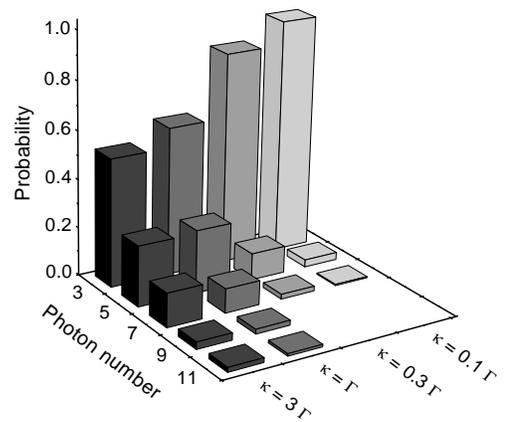,width=0.75\hsize}}\par\vspace{5mm}
\caption{Distribution of the number of photons escaping during
a single atomic transit for different cavity decay rates.
Otherwise, the parameters are the same as in Fig.~\ref{Fig:GHZ}.
For faster cavity decay, the probability for higher photon counts 
grows due to diabatic transitions within the dark state manifold.}
\label{Fig:nkappa}}
\end{figure}
%%%%%%%%%%%%%%%%%%%% Fig 18 %%%%%%%%%%%%%%%%%%%%%%%%%%%%%%%%%%%%%%%

An important question is the dependence of the triple correlations on the
cavity decay time. Simulating the adiabtic passage for different $\kappa$, we
found that the appearance of perfect correlations persisted even if $\kappa$
was increased to a value of $10\Gamma$, as long as only trajectories with
exactly three decay photons were considered. However, the number of
trajectories with five, seven, or more photons generated by diabatic transitions
during the atom's passage was found to grow with $\kappa$. This is illustrated
in Fig.~\ref{Fig:nkappa}, showing the distribution of the number of escaped
photons for four different cavity decay rates. Obviously, with more than three
photons generated in the cavity, no triple correlations can be expected. While
it is no problem to discard these trajectories in the numerical simulations, in
an actual experiment these processes may lead to wrong results due to finite
detection efficiency. If, for example, two out of five emitted photons are not
detected, the remaining photons are mistakenly registered as a three-photon
event but would show no definite correlations. The probability for these events
must therefore be minimized, which can be achieved by using a suitable detuning
and strength of the atom-cavity coupling to suppress diabatic transitions
during the passage of the atom (as described in Secs.~\ref{Sec:States}
and~\ref{Sec:QMC}) and by chosing a small cavity decay rate. Since excited
atomic states are occupied with an acceptably small probability only, the
system is not compromised by atomic spontaneous emission, which otherwise would
also result in the generation of excess photons.

%%%%%%%%%%%%%%%%%%%% Fig 19 %%%%%%%%%%%%%%%%%%%%%%%%%%%%%%%%%%%%%%%
\begin{figure}[tb]
\vbox{
\centerline{\psfig{file=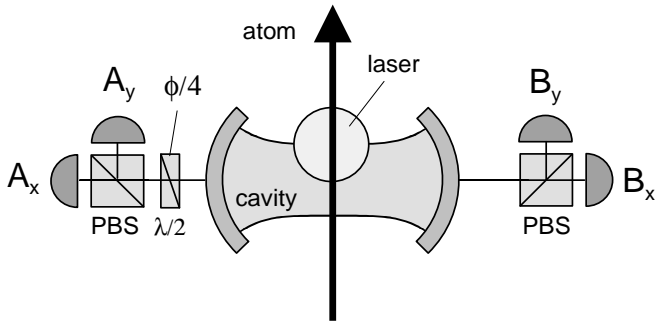,width=\hsize}}\par\vspace{5mm}
\caption{Simplified setup for a realistic experiment to probe
GHZ correlations of photons emitted from a cavity with two
output ports. The pair of detectors A analyzes the photon 
polarization in a rotated linear basis (given by the orientation 
of the wave plate), while detectors  B use a fixed linear 
basis. The GHZ argument may be tested by observing the result of 
analyzer B if both detectors {\bf\sf A$_x$} and {\bf\sf A$_y$}
have registered a photon.}
\label{Fig:GHZreal}}
\end{figure}
%%%%%%%%%%%%%%%%%%%% Fig 19 %%%%%%%%%%%%%%%%%%%%%%%%%%%%%%%%%%%%%%%

In an actual experiment, one may take advantage of some simplifictions. For the
experimental verification of the GHZ argument [cf. Eq.~(\ref{violation})], two
of the three analyzers should be set at the same variable angle
($\phi_1=\phi_2=\phi$), while the other is always set at an angle $\phi_3=0$.
Therefore, it is sufficient to use a pair of analyzers, with a half-wave plate
only in front of one of them. Instead of using beamsplitters to route the decay
photons, a cavity with two semitransparent output mirrors may be used with an
analyzer placed on either side. The proposed setup is shown in
Fig.~\ref{Fig:GHZreal}. To avoid problems due to detector dead time, events
should only be considered if all the photons from the passage of a single atom
are registered by different photon counters. If the probability for the
generation of more than three photons is negligible (which is the case, for
example, for a detuning $\delta\approx0.6\Gamma$ and a small cavity decay rate
$\kappa\approx0.1\Gamma$, cf. Figs.~\ref{Fig:gdetune} and~\ref{Fig:nkappa}),
this last condition may be fulfilled by discarding runs with less than three
distinct counters clicking.

In this scheme, GHZ correlations could be detected by examining the result of
the {\it B} counters when both {\it A} counters detect a photon. According to
Eq.~(\ref{GHZop}), the expectation value for the outcome of the {\it B}     
measurement
is then $-\cos(2\phi)$, as the product of the {\it A} results is always negative.
Perfect correlations result for $\phi=\pi/2$, when only the {\it B}$_x$-counter
should receive the third photon and for $\phi=0$, when the third photon should
always be registered by  {\it B}$_y$. To maximize the probability for two photons
being emitted through the mirror on the  {\it A}-side, and one through the  {\it B} mirror,
the  {\it A} mirror transmission $T$ should be twice that of the {\it B}-mirror. In this
case, the measurement record would show three separate photoelectric events in
the required combination of analyzers in 44.4\% of the atomic transits.

\section{Atom-photon GHZ correlations}
\label{Sec:GHZAtom}

The method proposed in Sec.~\ref{Sec:GHZprep} for the synthesis of
polarization-entangled states of the cavity field requires the initial
generation of coherent superpositions of Zeeman ground-state sublevels. While
certainly feasible, this requirement adds the need for additional microwave or
laser pulse manipulation of the atom before the adiabatic passage. The state
synthesis procedure would be greatly simplified if it were possible to generate
entangled states starting from a single Zeeman sublevel, which is easily
prepared by optical pumping. In this final section we will show that, by using
a different Zeeman scheme, it is indeed possible to generate a GHZ-type state
with an atom initially pumped to the $m_g=0$ ground state. However, in this
case one of the entangled particles is the atom itself, which must be detected
in order to observe GHZ correlations.

%%%%%%%%%%%%%%%%%%%% Fig 20 %%%%%%%%%%%%%%%%%%%%%%%%%%%%%%%%%%%%%%%
\begin{figure}[tb]
\vbox{
\centerline{\psfig{file=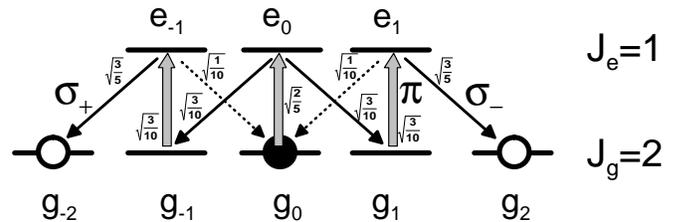,width=\hsize}}\par\vspace{5mm}
\caption{Level scheme of the $(J_g=2) \rightarrow (J_e=1)$ transition,
showing the transitions relevant for an atom prepared 
in the $m_g=0$ Zeeman sublevel (solid circle).
The wide arrows indicate the $\pi$-polarized pump transitions. 
The predominant cavity-induced transitions are shown as
solid arrows, the undesired transitions are dotted.
After the adiabatic passage, the atom is found in a
superposition of the $m_g=-2$ and the $m_g=+2$-level (open circles),
while the cavity contains two photons in a superposition of
$\sigma_+$  and $\sigma_-$ polarization.}
\label{Fig:arscheme}}
\end{figure}
%%%%%%%%%%%%%%%%%%%% Fig 20 %%%%%%%%%%%%%%%%%%%%%%%%%%%%%%%%%%%%%%%

The atomic detection should be efficient, have a good time resolution, and a low
background. These requirements are met by using metastable atoms, which can be
readily detected with secondary electron multipliers \cite{Fauls92,Obert96}. A
suitable transition is $1s_5\; (J=2) \rightarrow 2p_{10}\;(J=1)$ in argon,
starting from the metastable state $1s_5$, with a transition wavelength of
$912.30$~nm and a decay rate $\Gamma/2\pi=3.0$ MHz. Figure~\ref{Fig:arscheme}
shows the corresponding $J=2\rightarrow J=1$ level scheme. The adiabatic passage
procedure again uses a $\pi$-polarized pump field. The atom in this case must
be prepared in the $m=0$ level, e.g., by pumping on the $1s_5\; (J=2)
\rightarrow 2p_{8}\;(J=2)$ transition, and the cavity modes are assumed to be
empty. The lowest state of the dark space manifold that connects to this
initial state is
\begin{eqnarray}
\label{inversepassage}
|E_{0}\rangle &=&
-\sqrt{12} g^2(t)	|g_{0},0,0\rangle \nonumber \\
&&+ 2 g(t) \Omega(t)\left( |g_{-1},1,0\rangle + |g_{1},0,1\rangle\right)\\
&&- \Omega^2(t) \left( |g_{-2},2,0\rangle + |g_{2},0,2\rangle\right).
\nonumber
\end{eqnarray}
Due to the symmetry of the interaction Hamiltonian of Eq.~(\ref{Hamilt}) with
respect to $\sigma_+/\sigma_-$ and positive/negative magnetic quantum numbers
$m$, the adiabatic passage proceeds from the initial state to superposition
states of positive Zeeman levels with $\sigma_-$ photons and negative Zeeman
levels with $\sigma_+$ photons. In contrast to the case considered in
Sec.~\ref{Sec:GHZprep}, the adiabatic passage leaves the atom in a state
entangled with the cavity field:
\begin{equation}
|g_{0}\rangle \otimes |0,0\rangle
\Longrightarrow 
|g_{-2},2,0\rangle + |g_2,0,2\rangle.
\label{atomfinal}
\end{equation}

%%%%%%%%%%%%%%%%%%%% Fig 21 %%%%%%%%%%%%%%%%%%%%%%%%%%%%%%%%%%%%%%%
\begin{figure}[tb]
\vbox{
\centerline{\psfig{file=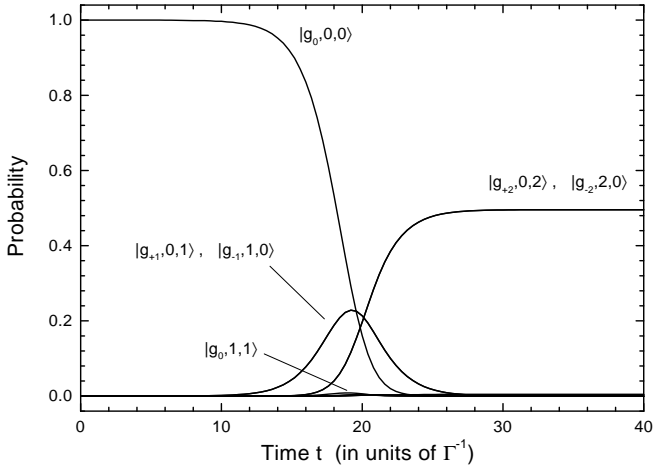,width=\hsize}}\par\vspace{5mm}
\caption{Monte Carlo wave-function simulation of 
two-po\-lari\-zation adiabatic passage in the case of a $J=2$ 
$\rightarrow$ $J=1$
transition. Atomic spontaneous emission is included, but cavity
decay is neglected. The system evolution is given in terms of the
occupation probability of the relevant basis states. Note that due to 
symmetry,
the curves for the levels $|g_m,0,m\rangle$ and $|g_{-m},m,0\rangle$
fall exactly on top of each other. 
The amplitudes of the Gaussian pulses for metastable argon
were chosen as $g_0=30\Gamma$ 
(at $t=17\Gamma^{-1}$) and $\Omega_0=50\Gamma$ (at $t=23\Gamma^{-1}$),
with FWHM $10\Gamma^{-1}$. The cavity detuning is $\delta=0.6\Gamma$.}
\label{Fig:argonsim}}
\end{figure}
%%%%%%%%%%%%%%%%%%%% Fig 21 %%%%%%%%%%%%%%%%%%%%%%%%%%%%%%%%%%%%%%%

As in Sec. \ref{Sec:QMC}, diabatic transitions may be suppressed and the system
dynamics restricted to adiabatic changes of state (\ref{inversepassage}) by
using a finite cavity detuning $\delta$. This is demonstrated by the results of
a Monte Carlo wave-function simulation of the system using the parameters for
the above specified transition in argon. In Fig.~\ref{Fig:argonsim}, the
occupation that is initially in level $|g_0,0,0\rangle$ is transferred with
equal probability to the levels $|g_1,1,0\rangle,|g_{-1},0,1\rangle$ to end
finally in a superposition of the states $|g_2,2,0\rangle,|g_{-2},0,2\rangle$,
where for these calculations cavity decay has been neglected.

%%%%%%%%%%%%%%%%%%%% Fig 22 %%%%%%%%%%%%%%%%%%%%%%%%%%%%%%%%%%%%%%%
\begin{figure}[t]
\vbox{
\centerline{\psfig{file=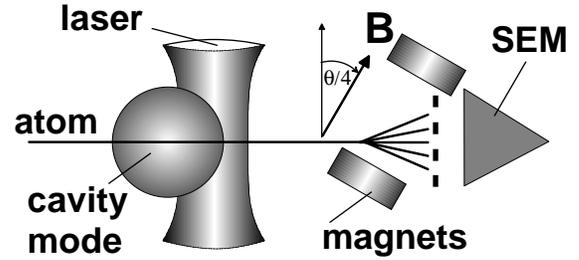,width=0.85\hsize}}\par\vspace{5mm}
\caption{Experimental setup for the detection of GHZ cor\-relations 
between two cavity decay photons and the atomic state.
The photons are detected as illustrated in Fig.~\ref{Fig:GHZreal}.
After the adiabatic passage, the atoms traverse a 
Stern-Gerlach detector with a magnetic field rotated by an angle
$\theta/4$. A grating then selects either the odd or the even orders
of the Zeeman-split atomic states, which are then detected by
a secondary electron multiplier (SEM).}
\label{Fig:GHZatomsetup}}
\end{figure}
%%%%%%%%%%%%%%%%%%%% Fig 22 %%%%%%%%%%%%%%%%%%%%%%%%%%%%%%%%%%%%%%%

The final state in Eq.~(\ref{atomfinal}) should also display the correlations
discussed by Greenberger, Horne, and Zeilinger, which are independent of the
type of particle considered. The cavity decay photons are detected by using
analyzers in front of the two cavity output ports as depicted in
Fig.~\ref{Fig:GHZreal}. The detection of the state-information carried by the
atom requires a different strategy. By analogy with the optical case, it should
involve the interference of positive and negative Zeeman states and allow for a
variable phase shift between both components. A possible experimental
realization of such a detector is a Stern-Gerlach analyzer with the magnetic
field oriented perpendicular to the cavity axis ($z$ axis), which is the
spatial quantization axis. The required phase shift is then obtained by
rotating the magnetic field axis in the $x$-$y$ plane by an angle $\theta/4$. A
sketch of the proposed setup is shown in Fig.~\ref{Fig:GHZatomsetup}. The
energy eigenstates in the analyzer are obtained from the following operator
involving angular momenta $J_\pm$ acting on the ground state
\begin{equation}
J(\theta)={1\over 2}( e^{-i\theta/4} J_+ + e^{i\theta/4} J_-),
\end{equation}
with eigenvalues $m_J(\theta)=0,\pm1,\pm2$. To establish GHZ correlations
between the atom and the photons emitted from the cavity, one considers the
operator with an eigenvalue of $1$ if the outcome of the Stern-Gerlach
measurement is even [$m_J(\theta)=0,\pm2$] and an eigenvalue of $-1$ if it is
odd [$m_J(\theta)=\pm1$]. This operator may be expressed as the difference of
the projectors onto even and odd eigenstates of $J(\theta)$:
\begin{eqnarray}
\label{GHZatomobserv}
M(\theta) &=& P_{\text{even}}(\theta) - P_{\text{odd}}(\theta) = \nonumber \\
&=&e^{i\theta}\; |g_{-2}\rangle\langle g_2 |  \; + \;
e^{i\theta/2}\; |g_{-1}\rangle\langle g_1|    \; + \;
|g_{0}\rangle\langle g_0 |    \nonumber
\\
&& + \;\;
e^{-i\theta/2}\; |g_{1}\rangle\langle g_{-1}| \; + \;
e^{-i\theta}\; |g_{2}\rangle\langle g_{-2}|. 
\end{eqnarray}
Note that the structure of $M(\theta)$ is entirely equivalent to that of
$L(\phi)$ defined in Eq.~(\ref{Ldef}), with the polarization-flip operators
$L_\pm$ replaced by operators $|g_{-m}\rangle\langle g_m|$ changing the sign of
a Zeeman substate. From Eq.~(\ref{GHZobserv}), the expectation value for triple
correlations of the two cavity photons (using the results of
Sec.~\ref{Sec:GHZdet}) and the atom may be calculated, assuming the system is
in state $|\text{GHZ}^\prime\rangle \equiv (|g_{-2},2,0\rangle +
|g_2,0,2\rangle)/\sqrt{2}$,
\begin{equation}
\label{GHZatomop}
\langle \text{GHZ}^\prime| L_1(\phi_1) L_2(\phi_2) M(\theta) |\text{GHZ}^\prime\rangle 
= \cos\left({\phi_1+\phi_2+\theta}\right).
\end{equation}
Thus, apart from the different origin of the third angle, Eq.~(\ref{GHZatomop})
is exactly analogous to Eq.~(\ref{GHZop}). Therefore, the conclusions of
Sec.~\ref{Sec:GHZdet} apply, in particular the discussion of perfect
correlations, which occur when ${\phi_1+\phi_2+\theta}$ is an integer multiple
of $\pi$.

%%%%%%%%%%%%%%%%%%%% Fig 23 %%%%%%%%%%%%%%%%%%%%%%%%%%%%%%%%%%%%%%%
\begin{figure}[tbp]
\vbox{
\centerline{\psfig{file=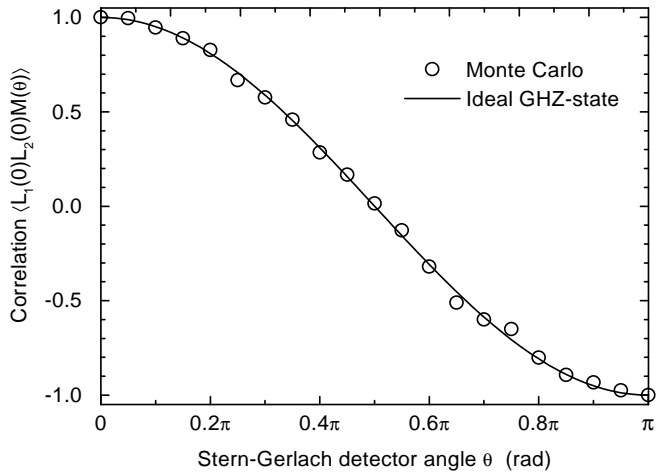,width=\hsize}}\par\vspace{5mm}
\caption{Monte Carlo wave-function calculation of averaged triple 
correlations between two cavity decay photons and the atomic 
state after the adiabatic passage. The parameters 
are as in Fig.~\ref{Fig:argonsim} and the cavity decay rate 
is $\kappa=0.1\Gamma$. 
The two photon analyzers are set at an angle $\phi_1=\phi_2=0$
and only the angle of the magnetic field $\theta$ in
the Stern-Gerlach detector is varied.
The solid line represents the correlations obtained in the
stationary state $(|g_{-2},2,0\rangle + |g_2,0,2\rangle)/\sqrt{2}$.}
\label{Fig:GHZatomcorr}}
\end{figure}
%%%%%%%%%%%%%%%%%%%% Fig 23 %%%%%%%%%%%%%%%%%%%%%%%%%%%%%%%%%%%%%%%

Results for the actual experimental situation, with photons decaying from the
cavity field before the adiabatic passage is complete, must again be obtained
by a Monte Carlo wave-function simulation of the full master equation. In
addition to detecting the quantum jumps of the cavity field, the atomic ground
state is probed at the end of each trajectory. An outcome for the measurement
is chosen randomly, weighted with probabilities obtained from the final
wave function, and the result is added to the classical record. Averaged triple
correlations obtained as a function of the orientation of the magnetic
field vector in the Stern-Gerlach detector are shown in
Fig.~\ref{Fig:GHZatomcorr}. The cosine function of Eq.~(\ref{GHZatomop}) is
well reproduced, showing that the dynamic generation of GHZ correlations is
also effective if the atomic degrees of freedom are entangled with the photons.

From an experimental point of view, GHZ correlations between an atom and two
photons may be easier to detect than their three-photon counterpart. With only
two photons escaping from the cavity, the fraction of usable detector responses
would increase to~50\%. In addition, detection of the atom provides a trigger
signal that can be used to discriminate against background noise in the photon
counters. Clearly, another advantage is that there is no need for the
preparation of a superposition of Zeeman ground states prior to the adiabatic
passage.

\section{Conclusions}
\label{Sec:Concl}

We have considered a model of adiabatic passage in a strongly coupled
atom-cavity system exposed to a classical pump field. With such a setup one may
synthesize nonclassical photon states by mapping Zeeman ground-state coherences
onto the cavity mode. The original proposal presented in
Refs.~\cite{Parki93,Parki95} included only a single cavity mode. We have
extended that scheme to include two degenerate cavity modes with orthogonal
polarization, the case relevant to Fabry-P{\'e}rot resonators used in the
laboratory. By analytical calculations as well as quantum Monte Carlo
simulations we have shown that atomic passage through the system does not lead
to pure number states of the two-mode cavity field, as long as atom, cavity,
and external pump field are resonant with each other. We could identify
diabatic transitions to degenerate dressed levels with higher photon number as
the source of this effect.

A central result of the paper is that the adiabatic passage scheme can be
modified to work even in the two-polarization case by introducing a finite
cavity detuning. This way diabatic transitions are suppressed and states with a
definite photon number of the cavity field may be efficiently generated.
Several applications of the method were discussed. As a first test of our
scheme, we calculated the synthesis of a three photon Fock state of a single
polarization of the cavity field and obtained an efficiency above 99\% based
on parameters actually achieved in cavity-QED experiments.

The system under study offers intriguing possibilities beyond the generation of
number states and their superposition for a single mode. An important property
of the atomic transition scheme we employ is its symmetry under a change of
sign of (circular) polarization and magnetic quantum number. Taking advantage
of this feature allows one to create polarization entangled states of the
cavity field. As a particularly relevant example we presented a proposal for
the synthesis of a so-called GHZ state consisting of three photons maximally
entangled in polarization. It can be used to test EPR correlations without
having to resort to Bell-type inequalities.

To probe correlations between the photons generated by our scheme, we make use
of the inevitable presence of cavity dissipation. The decay photons emitted
from the cavity are sent to linear polarization analyzers of variable
orientation. For certain combinations of detector orientations, perfect
correlations between the photons may be observed. Most remarkably, these
correlations persist even if cavity decay and state synthesis occur on the same
time-scale.

One should note a difference between the states synthesized in the cavity and
the states discussed by GHZ \cite{GHZ89}. In the latter case, each particle
occupies a distinct mode, while in our cavity the photons are, in principle,
not distinguishable. Nonetheless, as demonstrated in Figs.~\ref{Fig:corr}
and~\ref{Fig:GHZatomcorr}, the observed correlations for a particular
subensemble would be precisely as for the ``ideal" GHZ state. Qualitatively,
this correspondence is due to the temporal evolution of the adiabatic passage
and the stochastic nature of cavity decay, with photons usually emerging from
the resonator ``one by one." This effect could be enhanced through manipulation
of the adiabatic passage process: combining suitable shifts of the atomic
states with multipeaked interaction profiles $g(t)$ and $\Omega(t)$, the
adiabatic passage would proceed in well-separated stages with exactly one
photon emitted per stage. With the photons arriving at the detectors at
distinct (predetermined) times, the observed correlations generated by the
adiabatic passage should correspond exactly to those predicted by GHZ without
the need for post selection of a particular subensemble.

A practical concern for the current scheme is the limited time resolution of
the detectors, which may not be sufficient to separate the decay events. In
this case one might simply register only events in which each of the three
photons is actually detected by a separate detector. Note that, in the three
particle interferometer orginally proposed by GHZ \cite{Green90}, the correct
direction and energy of the particles must also be enforced by apertures around
the source and filters in front of the detectors.

We did not discuss the influence of finite detection efficiency on the
correlations observed. However, as long as the number of photons generated does
not exceed three (a condition that, as we have demonstrated, may be fulfilled
by chosing a sufficiently large cavity detuning or coherent coupling $g_0$),
missing a photon decaying from the cavity would only lower the coincidence rate
observed, but would not lead to modified three photon correlations. The reason
is that events with less than three detected photons can easily be discarded in
a coincidence measurement. This argument is sometimes called the {\em fair
sampling assumption} \cite{Claus69}.

The proposed method for the synthesis of entangled states is not restricted to
the specific cases discussed in the paper. More complicated entangled states of
the cavity field may be generated as well. Examples include superposition
states with different photon number in the two polarization modes (e.g.,
$|3,0\rangle + |0,2\rangle$) or entangled states with polarization distributed
in a more general way (e.g.,  $|4,2\rangle + |1,3\rangle$), as would be useful
in quantum information networks with error correction \cite{Cirac97}.

Maximally entangled states with more than three particles, as discussed by
Mermin \cite{Mermi90b}, are of particular interest, since they provide an
exponentially stronger test of local realism. These states may be generated
with our scheme without any modification, if an atomic transition with the
corresponding angular momentum quantum number is used. In cesium, a four
particle generalization of the GHZ state may be synthesized with the
$F_g=4\rightarrow F_e=4$ hyperfine component of the D2 line. However, with
increasing number of particles, detection of the correlations between them
becomes considerably more challenging, since the coincidence rate would be
severely reduced.

The possibilities for the generation of entanglement in our system are further
expanded by including atom-photon correlations in the considerations. A
corresponding experiment would offer at least two additional benefits: instead
of a coherent superposition of Zeeman sublevels, only a single atomic substate
is needed as an initial condition for the adiabatic passage, so that the
preparation of the atomic state would require only standard optical pumping.
Furthermore, detection of the atom would provide a gate for the photon
detectors, increasing the signal to noise ratio.

In this paper we have restricted our attention to the case of a dilute atomic
beam passing through a cavity. In view of experimental progress in the field
\cite{Mabuc96,Hood98}, one may also consider atoms moving through the cavity at
a low enough speed to be considered stationary on the timescale set by the
coherent interaction, or even trapped atoms or ions in a cavity. The method
presented here could also be applied to the case of stationary atoms in a
cavity, if the dynamics that in ordinary adiabatic passage is induced by the
atom's motion, is generated instead by suitably timed pulses. The interaction
with the two cavity modes may be controlled by coupling the atomic levels with
a Raman transition using the cavity field and an auxiliary classical field. As
suggested in Refs.~\cite{Law96a,Law97,Cirac97}, the Raman coupling strength may
then be adjusted by changing the intensity of the auxiliary field.
 
\acknowledgments
The authors wish to thank Hideo Mabuchi, Sze Men Tan, and Peter Zoller for
valuable discussions, and S.M.T. especially for his assistance with setting up the
Monte Carlo simulations. This work was supported by the Defense Advanced
Projects Agency via the initiative for Quantum Information and Computation
(QUIC) administered by the Army Research Office, by the National Science
Foundation, and by the Office of Naval Research. W.L. also gratefully
acknowledges support from the Deutsche Forschungsgemeinschaft.

\nocite{Bragi89}

\end{document}